\documentclass[12pt,epsf,qsymbols]{article}
\usepackage[utf8]{inputenc}
\usepackage[T1]{fontenc}
\usepackage[normalem]{ulem}
\usepackage[french,croatian,serbian,english]{babel}
\usepackage{tabularx}
\usepackage{array}
\usepackage{graphics}
\usepackage[pdftex]{graphicx}
\usepackage{psfrag}
\usepackage{epsfig}
\usepackage{subfig}
\usepackage{mathtools}
\usepackage{amssymb}
\usepackage{bbold}
\usepackage{setspace}
\usepackage{rotating}
\usepackage{colortbl}
\usepackage{longtable}
\usepackage{slashed}
\usepackage{braket}
\usepackage{lineno}
\makeatletter
\usepackage{textcomp}
\usepackage[usenames,dvipsnames]{xcolor}
\usepackage{relsize}
\usepackage{cite}
\usepackage{slashed}
\usepackage{float}
\usepackage{cancel}
\usepackage{bbold}
\usepackage{bbm}
\usepackage{bm}
\usepackage{enumerate}
\usepackage{amsmath}
\usepackage{multirow}
\usepackage{verbatim}
\usepackage{placeins}

\usepackage{hyperref}
\hypersetup{colorlinks,citecolor=nicegreen,linkcolor=niceblue}
\hypersetup{colorlinks=true}

\usepackage{pifont}

\allowdisplaybreaks

\setlongtables

\newcommand{\pdir}{p\kern -5.2pt\raise 0.2ex\hbox {/}}

\newcommand{\vdir}{v\kern -5.75pt\raise 0.15ex\hbox {/}}
\newcommand{\kdir}{k\kern -5.75pt\raise 0.15ex\hbox {/}}
\newcommand{\epsdir}{\epsilon\kern -5.0pt\raise 0.15ex\hbox {/}}
\newcommand{\bvdir}{\bar{v}\kern -5.75pt\raise 0.15ex\hbox {/}}
\newcommand{\Ddir}{D\kern -7.75pt\raise 0.20ex\hbox {/}}
\newcommand{\Adir}{A\kern -7.75pt\raise 0.20ex\hbox {/}}
\newcommand{\ldir}{l\kern -5.0pt\raise 0.2ex\hbox{/}}
\newcommand{\varepsdir}{\varepsilon\kern -5.5pt\raise 0.15ex\hbox{/}}

\makeatother

\definecolor{niceblue}{rgb}{0.15,0.15,0.6}
\definecolor{nicegreen}{rgb}{0.1,0.5,0.1}
\definecolor{Red}{rgb}{1.,0.,0.}

\definecolor{Green}{rgb}{0.2,.7,0.2}

\begin{document}
\unitlength = 1mm

\thispagestyle{empty} 
\begin{flushright}
\begin{tabular}{l}
\end{tabular}
\end{flushright}
\begin{center}
\vskip 3.4cm\par
{\par\centering \textbf{\LARGE  
\Large \bf Integrating Out New Fermions at One Loop }}
\vskip 1.2cm\par
{\scalebox{.85}{\par\centering \large  
\sc Andrei Angelescu$^a$, Peisi Huang$^a$}
{\par\centering \vskip 0.7 cm\par}
{\sl 
$^a$~{Department of Physics and Astronomy\\ University of Nebraska-Lincoln, Lincoln, NE, 68588, USA.}}\\

{\vskip 1.65cm\par}}
\end{center}

\vskip 0.85cm
\begin{abstract}

We present the fermionic universal one--loop effective action obtained by integrating out heavy vector--like fermions at one loop using functional techniques. Even though previous approaches are able to handle integrating out heavy fermions with non--chiral interactions, i.e. vanishing $\gamma^5$ interaction terms, the computations proceed in a tedious manner that obscures a physical interpretation. We show how directly tackling the fermionic functional determinant not only allows for a much simpler and transparent computation, but is also able to account for chiral interaction terms in a simple, algorithmic way. Finally, we apply the obtained results to integrate out at one loop the vector--like fermions appearing in a toy model and in a fermionic model that exhibits strong cosmological phase transitions.

\end{abstract}
\newpage
\setcounter{page}{1}
\setcounter{footnote}{0}
\setcounter{equation}{0}
\noindent

\renewcommand{\thefootnote}{\arabic{footnote}}

\setcounter{footnote}{0}

\tableofcontents

\newpage


\section{Introduction} \label{sec:intro}

Despite a huge experimental effort, signals of physics beyond the Standard Model (SM) remain elusive up to date in direct searches at the Large Hadron Collider (LHC). The absence of such signals has determined a shift of attention towards possible indirect effects of heavy new particles, which can be systematically studied in the context of effective field theories (EFTs) such as the Standard Model Effective Theory (SMEFT)~\cite{Weinberg:1980wa,Buchmuller:1985jz}. 

The main strength of EFTs is model independence: experimental measurements can be used to place bounds on the higher--dimensional deformations of a given low--energy theory (e.g. the SM) without specifying any underlying ultraviolet (UV) theory responsible for inducing those deformations. Conversely, the heavy degrees of freedom of a given UV theory can be integrated out and matched to a low--energy EFT, allowing for an efficient study of the resulting effects. In this sense, EFTs serve as a bridge between UV theories and experiments, as long as the new degrees of freedom are sufficiently heavier than the energy scale of the experiment.

Given the current (and future) experimental precision, an accurate translation of the SMEFT bounds to bounds on specific new physics scenarios requires the matching to be performed at the level of (at least) one loop. Traditionally, this task is done at the amplitude level with the help of Feynman (loop) diagrams. A perhaps more elegant and simpler alternative for performing the one--loop matching relies on working directly with the path integral. The idea behind this approach is to identify the contributions of the heavy fields to the one--loop functional determinant, and then expand the determinant in inverse powers of the heavy masses to obtain effective operators containing the light fields. Among the desirable features of such functional methods, there are at least two worth mentioning. Firstly, unlike in the case of Feynman diagrams, gauge covariance is preserved in the intermediate steps by performing a covariant derivative expansion (CDE), which automatically ensures a gauge--invariant final result. Secondly, such methods give universal results that can be applied to a broad class of new physics scenarios with almost no assumption regarding the UV dynamics.

Functional techinques for one--loop matching and their applications were first developed almost 40 years ago in Refs.~\cite{Gaillard:1985uh,Chan:1986jq,Cheyette:1987qz}. The subject has been recently brought back into focus by Ref.~\cite{Henning:2014wua}, which provided a universal master formula for one--loop matching assuming degenerate masses for the new particles. The generalization to non--degenerate spectra was completed in Ref.~\cite{Drozd:2015rsp}, and the resulting master formula was named the ``Universal One--Loop Effective Action'' (UOLEA). However, as pointed as out in Ref.~\cite{delAguila:2016zcb} (see also Ref.~\cite{Boggia:2016asg}), both master formulas presented in Refs.~\cite{Henning:2014wua,Drozd:2015rsp} could only account for the so--called ``heavy--only'' terms, i.e. terms stemming from loops containing only heavy fields. Shortly afterwards, master formulas including also the ``(mixed) heavy--light'' terms (originating from loops containing both light and heavy fields) were computed using various methods in Refs.~\cite{Henning:2016lyp,Ellis:2016enq,Fuentes-Martin:2016uol}. With the help of the covariant diagram technique developed in Ref.~\cite{Zhang:2016pja}, Ref.~\cite{Ellis:2017jns} established that heavy--only and heavy--light terms share the same structure, and extended the UOLEA of Ref.~\cite{Drozd:2015rsp} to include heavy--light terms as well. In addition to the references mentioned up to now, applications of functional methods for one--loop matching have been studied e.g. in Refs.~\cite{Dittmaier:1995cr,Dittmaier:1995ee,Chiang:2015ura,Huo:2015exa,Huo:2015nka,Wells:2017vla,Jiang:2018pbd} (see also Refs.~\cite{Haisch:2020ahr,Gherardi:2020det}).

Although some steps have been already taken in this direction in Ref.~\cite{Kramer:2019fwz}, the universal master formulas available to date in the literature do not systematically capture the effects of integrating out heavy fermions at one loop. The main reason behind this is the presence of fermionic interaction terms containing the $\gamma^5$ Dirac matrix. Furthermore, even if the terms containing $\gamma^5$ are set to zero, applying the existing master formulas to the fermionic case proves to be a tedious task.

It is therefore the aim of this paper to provide a universal master formula that is suitable for integrating out heavy fermions at one loop. To this end, we calculate for the first time the heavy--only contributions arising from integrating out heavy vector--like fermions (VLFs) at one--loop, consistently taking into account the effects of $\gamma^5$. We choose to perform this operation in the unbroken (symmetric) phase, where none of the light scalars have a vacuum expectation value (VEV). In the familiar case where the low--energy theory is the SM, this amounts to calculating the SMEFT dimension--6 operators and Wilson coefficients induced by loops containing only heavy VLFs.

Our work is outlined as follows. In Sec.~\ref{sec:setup}, we specify the general Lagrangian containing all the VLF interaction terms allowed in the unbroken phase, and then set up the expansion of the functional determinant in inverse powers of the VLF masses. We also provide a brief review of alternative methods of tackling the fermionic functional determinant. Sec.~\ref{sec:fermionic-uolea} is dedicated to the calculation of the fermionic UOLEA terms and their associated universal coefficients. The results from this section are connected to previous universal results in App.~\ref{app:uolea-coeffs}, and then summarized in App.~\ref{app:equal-masses}. In Sec.~\ref{sec:examples}, we apply the results obtained in Sec.~\ref{sec:fermionic-uolea} to a fermionic toy model and a more realistic VLF model. Finally, we conclude in Sec.~\ref{sec:conclusions}.

\section{Setup} \label{sec:setup}

We consider a fermionic model containing several heavy Dirac VLF multiplets $\Psi_i$, charged under a generic (semi--simple) gauge group, e.g. the SM gauge group $SU(3)_C \otimes SU(2)_L \otimes U(1)_Y$. Working in the unbroken phase of the theory, the most general renormalizable and gauge--invariant Lagrangian of the VLF sector reads
\begin{equation}
\label{eq:vlf_lagr}
\mathcal{L}_{\rm VLF} = \overline{\Psi} \left( i \gamma_\mu D^\mu - M - S - i \gamma^5 P \right) \Psi = \overline{\Psi}_i \left[ \left( i \gamma_\mu D^\mu_{i} - m_{i} \right) \delta_{ij} - S_{ij} - i \gamma^5 P_{ij} \right] \Psi_j,
\end{equation}
where summing over the multiplet (flavour) indices $i,j$ is implicit. Throughout this paper we use the latin indices $\{i,j,k,l,m,n\}$ to denote flavour indices and assume that summing over them is implicit when considering Lagrangians.~\footnote{We will however assume that summing over flavour indices is not performed in the case of definitions/expressions of the universal coefficients, which appear later on.} By virtue of working in the unbroken phase, the vector--like mass and heavy fermion covariant derivative matrices are diagonal in multiplet space, $ M_{ij} = m_{i} \delta_{ij} $ and $D^\mu_{ij} = D^\mu_{i} \delta_{ij} $  (no sum), whereas the $S$ and $P$ matrices are in general non--diagonal, which implies that they do not always commute with $M$. Hermicity of the Lagrangian in Eq.~\eqref{eq:vlf_lagr} implies that both $S$ and $P$ are Hermitian. 

The interactions of the heavy fermion fields $\Psi_{i}$ with the light gauge fields are encoded in the $D_\mu$ matrix, which takes a generic form
\begin{equation}
\label{eq:covar-deriv-def}
D_{\mu,i} = \partial_\mu + i \, \sum g \, G_\mu^a \, t^a_{R,i},
\end{equation}
where the sum runs over all the light gauge fields $G$, and the repeated index $a$ implies summing over all generators of each gauge group. The index $i$ of the group generator $t^a_{R,i}$ serves as a reminder that it lies in the same gauge group representation as the heavy fermion $\Psi_i$ that it acts upon. On the other hand, $S = S(\phi)$ and $P = P(\phi)$ specify the Yukawa interactions of $\Psi_i$ with light (pseudo)scalar fields denoted generically by $\phi$.

Since we are working in the unbroken phase, where none of the scalars appearing in $S$ and $P$ have a vacuum expectation value (VEV), the vector--like nature of the new fermions $\Psi$ forbids an axial term $ \overline{\Psi}_i \left( \gamma_\mu \gamma^5 A_{ij}^\mu \right) \Psi_j$, while we discard the tensor term $ \overline{\Psi}_i \left( \sigma_{\mu\nu} T_{ij}^{\mu\nu} \right) \Psi_j$, as it encodes non--renormalizable interaction terms of at least dimension--5. The pseudoscalar term $i \gamma^5 P_{ij}$ is however allowed, as the left and right chiralities of a given VLF can couple differently to the scalar fields present in $S$ and $P$. If we were to work in the broken phase, the general Lagragian would contain both the axial term mentiond above and a non--diagonal vector term, $ \overline{\Psi}_i \left( \gamma_\mu V_{ij}^\mu \right) \Psi_j$.

To obtain the effective action induced by integrating out the heavy fermions $\Psi_i$ at one loop, we must calculate the ``tracelog''~\cite{Gaillard:1985uh,Chan:1986jq,Cheyette:1987qz,Henning:2014wua} of the operator appearing in Eq.~\eqref{eq:vlf_lagr}:
\begin{equation}
\label{eq:tr-log}
S_H = i \, c_f \, {\rm Tr} \log \left( i \gamma_\mu D^\mu - M - S - i \gamma^5 P \right).
\end{equation}
Here, the fermionic factor $c_f = -1$ appears from performing a gaussian path integral over anti--commuting fermionic fields. The symbol $\rm Tr$ denotes a full trace over coordinate space, flavour space, and internal degrees of freedom (spin and gauge). Computing the effective action up to dimension--$n$ amounts to expanding $S_H$ up to terms of $\mathcal{O}(M^{4-n})$. To set up this expansion, the usual procedure is to write explicitly the trace over coordinate space using momentum eigenstates in $d$ dimensions:
\begin{align}
\label{eq:tr-log-2}
S_H &= i \, c_f \, {\rm tr} \int \frac{d^d p}{(2\pi)^d} \langle p | \log \left( i \gamma_\mu D^\mu - M - S - i \gamma^5 P  \right) | p \rangle \notag \\
& =  i \, c_f \, {\rm tr} \int d^d x \int \frac{d^d p}{(2\pi)^d} e^{-ipx} \log \left( i \gamma_\mu D^\mu - M - S - i \gamma^5 P  \right) e^{ipx} \notag \\
& = i \, c_f \, {\rm tr} \int d^d x \int \frac{d^d p}{(2\pi)^d} \log \left(- \slashed{p} - M + i \gamma_\mu D^\mu - S - i \gamma^5 P  \right),
\end{align}		
where for the last equality we used the fact that ``sandwiching'' an operator function $f(\partial_\mu)$ between the exponential amounts to shifting $f(\partial_\mu) \to f(\partial_\mu + i p_\mu)$. At this stage, one can ``sandwich'' the $\log$ in Eq.~\eqref{eq:tr-log-2} by $\exp \left( \pm i D_\mu \frac{\partial}{\partial p_\mu} \right)$, as originally done in Refs.~\cite{Gaillard:1985uh,Cheyette:1987qz}, and obtain an expansion that is manifestly gauge--invariant at all intermediate stages~\cite{Henning:2014wua}, i.e. covariant derivatives appear only in commutators. We do not pursue this avenue in this paper and instead work directly with the expression from Eq.~\eqref{eq:tr-log-2}.

The trace operator $\rm tr$ in Eq.~\eqref{eq:tr-log-2} denotes tracing over flavour, spin, and gauge degrees of freedom. After flipping the sign of the loop momentum, we get:
\begin{align}
\label{eq:tr-log-3}
S_H &= i \, c_f \, {\rm tr} \int d^d x \int \frac{d^d p}{(2\pi)^d} \log \left[\slashed{p} - M -  \left( -i \gamma_\mu D^\mu + S + i \gamma^5 P \right)  \right] \notag \\
&= - \frac{c_f}{16 \pi^2} {\rm tr} \int d^d x \int [d^d p] \log \left[1 -  \left( \slashed{p} - M \right)^{-1}  \left( -i \gamma_\mu D^\mu + S + i \gamma^5 P \right)  \right].
\end{align}
Passing to the second line of Eq.~\eqref{eq:tr-log-3} is possible only because the of the equality $ {\rm tr} \log(A+B) = {\rm tr} \log(A) + {\rm tr} \log(1 + A^{-1}B)$. Moreover, we have discarded the constant term $\propto \log(\slashed{p}-M)$ as it does not depend on any fields, and defined a new momentum integral measure as:
\begin{equation*}
\int \frac{d^d p}{(2\pi)^d} \equiv \frac{i}{16 \pi^2} \int [d^d p].
\end{equation*}
Denoting the fermionic propagator as $\left( \slashed{p} - M \right)^{-1} \equiv \slashed{\Delta} $, we can finally perform the Taylor expansion of the $\log$ from Eq.~\eqref{eq:tr-log-3} and write down the effective one--loop Lagrangian as
\begin{equation}
\label{eq:effective-lagrangian}
\mathcal{L}_H = \frac{c_f}{16 \pi^2} \sum_{n=1}^{\infty} \frac{1}{n}  \int [d^d p] \; {\rm tr} \left\{ \left[ \slashed{\Delta} \left(-i \gamma_\mu D^\mu + S + i \gamma^5 P \right)  \right]^n \right\}.
\end{equation}
A desirable feature of the expansion in Eq.~\eqref{eq:effective-lagrangian} is that each order $n$ contains only operators of dimension--$n$, which renders the power counting transparent. This owes to the fact that all three terms $D^\mu$, $S$, and $P$ are of dimension--1, as they depend linearly on bosonic fields. Therefore, truncating the series at $n=6$ ensures the inclusion of all the effective operators arising at dimension--6 or lower. 

In the existing literature, several methods have been advanced for evaluating the fermionic functional trace. Refs.~\cite{Henning:2014wua,Drozd:2015rsp,Henning:2016lyp,Fuentes-Martin:2016uol} use the invariance of the trace under $\gamma_\mu \to - \gamma_\mu$ and $ {\rm Tr} \log A + {\rm Tr} \log B = {\rm Tr} \log AB $ to bring the fermionic trace in a form that resembles the bosonic trace:
\begin{align}
\label{eq:squaring-the-determninant}
S_H^{\rm ferm} &= \frac{i c_f}{2} \left[  {\rm Tr} \log \left( i \gamma_\mu D^\mu - M -  W \right) + {\rm Tr} \log \left( -i \gamma_\mu D^\mu - M -  W \right) \right] \notag \\
& \equiv \frac{i c_f}{2}  {\rm Tr} \log \left( D^2 + M^2 + U_{\rm ferm}  \right),
\end{align}
with
\begin{equation}
\label{eq:U-ferm}
U_{\rm ferm} \equiv - \frac{i}{2} \sigma^{\mu\nu} [D_\mu,D_\nu] - i \gamma^\mu [D_\mu, W] - i  [\gamma^\mu, W] D_\mu + \{ M, W \} + W^2,
\end{equation}
where we use the usual notation for the (anti)commutator and $\sigma^{\mu\nu} = \frac{i}{2} [\gamma^\mu, \gamma^\nu]$. Once brought into the form shown in Eq.~\eqref{eq:squaring-the-determninant}, the fermionic trace can be computed using the known results for the bosonic trace~\cite{Henning:2014wua,Drozd:2015rsp}, dubbed the ``Universal One--Loop Effective Action'' (UOLEA). However, these results apply solely to the case where $U_{\rm ferm}$ does not contain any open covariant derivatives,~\footnote{As usually defined in the literature, open covariant derivatives are covariant derivatives that do not appear inside commutators.} which is true only if $[\gamma^\mu, W] = 0$. Since in the general renormalizable case displayed in Eq.~\eqref{eq:vlf_lagr} $W$ contains $\gamma^5$, it is clear that the trick shown in Eq.~\eqref{eq:squaring-the-determninant} is helpful only if the $\gamma^5$ piece is vanishing, i.e. $P=0$. And even if $P=0$ and therefore $[\gamma^\mu, W] = 0$, the rather lengthy expression of $U_{\rm ferm}$ from Eq.~\eqref{eq:U-ferm} makes this method impractical for computing the fermionic functional trace.

Another method to compute the fermionic functional trace was put forward in Ref.~\cite{Zhang:2016pja} and relies on a diagrammatic computation of Eq.~\eqref{eq:effective-lagrangian}. Exploiting the cyclic property~\footnote{Note that the cyclic property of the trace in Eq.~\eqref{eq:effective-lagrangian} is not apparent, as $D^\mu$ acts on coordinate space, while $\rm tr$ does not include a trace over coordinate space. This issue is discussed later on in Sec.~\ref{subsec:dim4}.} and the gauge invariance of the trace in Eq.~\eqref{eq:effective-lagrangian}, one can write down all the allowed operators appearing at a given order and then calculate the corresponding universal coefficients by considering loop diagrams involving combinations of insertions of the three terms from Eq.~\eqref{eq:effective-lagrangian} ($i \gamma_\mu D^\mu$, $ S $, and $i \gamma^5 P$). This method has been dubbed the ``covariant diagram'' approach, as the covariant derivative $D_\mu$ is treated as a single object, as opposed to being split into $\partial_\mu$ and the gauge boson piece, as is done in conventional Feynman diagrams. Our computation of the fermionic functional trace will be similar in spirit to the covariant diagram approach, as we also rely on the cyclic property of the trace. However, as opposed to Ref.~\cite{Zhang:2016pja}, we do not use diagrams, but read the terms directly from Eq.~\eqref{eq:effective-lagrangian}. Moreover, Ref.~\cite{Zhang:2016pja} splits the fermionic propagator into two parts:
\begin{equation*}
\frac{\gamma^\mu p_\mu}{p^2 - m^2} + \frac{m}{p^2 - m^2}.
\end{equation*}
In our approach, instead of performing this separation, we just compute the fermion traces and the resulting loop integrals using Package--X~\cite{Patel:2015tea,Patel:2016fam}.

Finally, we also mention Ref.~\cite{Kramer:2019fwz}, which discusses the fermionic extension of the UOLEA (both heavy--only and heavy--light contributions, plus mixed scalar--fermion contributions) and calculates all the relevant momentum integrals, but without specifying the $\gamma$--matrix structure of the interactions and therefore without evaluating the spin traces.

\section{Computation of the Fermionic One--Loop Effective Action} \label{sec:fermionic-uolea}

We now compute order--by--order the terms from the effective lagrangian in Eq.~\eqref{eq:effective-lagrangian}, up to $n=6$, which corresponds to dimension--6 operators. In the particular case of the SMEFT~\cite{Grzadkowski:2010es}, where the only light scalar field is the SM Higgs doublet, gauge invariance dictates that terms corresponding to odd powers of $n$ vanish. Nevertheless, we focus on the general case and consider odd powers of $n$ as well. Throughout the computation, we denote the trace over spin degrees of freedom (i.e. Dirac traces) as $\rm tr_s$, and the trace over gauge indices as $\rm tr_g$. Moreover, we pull out from each term in the effective action a factor of 
\begin{equation}
\label{eq:no-dof-dirac}
{\rm tr_s} \, {\mathbb{1}} \equiv n_D = 4,
\end{equation}
which represents the number of spin degrees of freedom for a Dirac fermion (we denote the identity matrix in spinor space as $\mathbb{1}$). This choice facilitates the comparison with the UOLEA results (see App.~\ref{app:uolea-coeffs}).  In addition, we explicitly write down the symmetry factors for each gauge--invariant trace, as opposed to absorbing them in the universal coefficients. For example, the ${\rm tr_g} \left( S_{ij} S_{jk} S_{ki} \right)$ term comes with a factor $\frac{1}{3}$ in front, whereas ${\rm tr_g} \left( S_{ij} P_{jk} P_{ki} \right)$ has no symmetry factor. For the universal coefficients, we use the shorthand notation
\begin{equation}
\label{eq:univ-coeff-notation}
g (m_i, m_j, m_k, m_l, \cdots) \equiv g^{ijkl \cdots}.
\end{equation}
Also, we often encounter coefficients with flipped signs for some of the masses. To this end, we define the following notation:
\begin{equation}
\label{eq:univ-coeff-notation-2}
g (m_i, m_j, -m_k, m_l, \cdots) \equiv g^{ij(k)l \cdots}, \quad g (m_i, -m_j, m_k, -m_l, \cdots) \equiv g^{i(j)k(l) \cdots},
\end{equation}  
i.e. an index between brackets translates to a flipped sign for the corresponding mass. As mentioned before, we consider the sum over flavour indices to be implicit.

Finally, before starting our computation, we briefly discuss the type of terms that are allowed in the one--loop effective action. Even if not apparent from Eq.~\eqref{eq:effective-lagrangian}, gauge invariance ensures that in the final result all the covariant derivative matrices organize into commutators,~\footnote{Even if not true in general, this statement holds for our particular case, where $S$ and $P$ do not contain any open covariant derivatives, i.e. covariant derivatives not appearing in commutators. For example, open covariant derivatives would be present if heavy scalars and/or gauge bosons would be integrated out alongside the VLFs.} i.e. the only dependence on covariant derivatives is through pieces such as $\left[ D_\mu,X \right]$ (with $X=S,P$) or $\left[D_\mu, D_\nu \right]$, which we denote as
\begin{equation}
\label{eq:F-munu}
\left[D_\mu, D_\nu \right] \equiv i F_{\mu\nu}.
\end{equation} 
Moreover, all terms containing an odd number of $D_\mu$'s vanish because of Lorentz invariance. Due to the properties of Dirac traces involving $\gamma^5$, the only terms containing odd powers of $P$ are $\mathcal{O}(P D^4)$ and  $\mathcal{O}(S P D^4)$, which appear at dimension--5 and dimension--6, respectively.

\subsection{Dimension--1 Terms} \label{subsec:dim1}

At dimension--1, there is only one possible term, $\mathcal{O}(S)$ , all the other ones being forbidden by gauge and Lorentz invariance:
\begin{equation}
\label{eq:lagr-n=1}
16 \pi^2 \mathcal{L}_H^{n=1} = c_f   \int [d^d p] \,  {\rm tr} \left[ \slashed{\Delta}_i  S_{ii}   \right] = c_f \, n_D \,  g_1^i \, {\rm tr_g} \left( S_{ii} \right),
\end{equation} 
where $g_1^i$ can be expressed in terms of the master integrals from App.~\ref{app:master-integrals} as
\begin{equation}
\label{eq:g1}
g_1^i = \frac{1}{n_D} \int [d^d p] \,  {\rm tr_s} \left[ \slashed{\Delta}_i \right]  = m_i \, \mathcal{I}[q^0]^i_1.
\end{equation}
The factor $n_D$ from Eq.~\eqref{eq:lagr-n=1} comes from the trivial spin trace $\rm tr_s \, \mathbb{1}$, and the terms involving $\gamma_\mu$ and/or $\gamma^5$ matrices vanish under the spin trace. Note that, by gauge invariance, the sole term from Eq.~\eqref{eq:lagr-n=1} vanishes unless the light particle spectrum contains a real singlet. In this case, this term would represent a tadpole term for the singlet.

\subsection{Dimension--2 Terms} \label{subsec:dim2}

At dimension--2, only the $\mathcal{O}(S^2)$ and $\mathcal{O}(P^2)$ terms appear, as the remaining $\mathcal{O}(D^2)$ and $\mathcal{O}(D_\mu S, D_\mu P)$ terms are forbidden by gauge and Lorentz invariance, respectively. Therefore, we can safely drop the $i \slashed{D}$ piece from the $n=2$ contribution:
\begin{align}
\label{eq:lagr-n=2}
16 \pi^2 \mathcal{L}_H^{n=2} &= \frac{1}{2} c_f \int [d^d p] \, {\rm tr} \left[  \slashed{\Delta}_i  \left(S_{ij} + i \gamma^5 P_{ij} \right) \slashed{\Delta}_j  \left(S_{ji} + i \gamma^5 P_{ji} \right)  \right] \notag \\
&= \frac{1}{2} c_f \, n_D \left[   g_2^{ij} \, {\rm tr_g} \left( S_{ij} S_{ji} \right) + \left( g_2^{i(j)} + \delta  g_2^{ij} \right) \, {\rm tr_g} \left( P_{ij} P_{ji} \right) \right],
\end{align}
where we have kept the symmetry factor $\frac{1}{2}$ of the gauge traces involved. Both terms quantify the one--loop contribution of the heavy fermions to the masses of the light scalars present in the theory. Using dimensional regularization~\cite{tHooft:1972tcz} in $d = 4 - 2\epsilon$ dimensions, the universal coefficients are given by:
\begin{align}
\label{eq:g2}
\begin{split}
g_2^{ij} &= \frac{1}{n_D} \int [d^d p]  \, {\rm tr_s} \left( \slashed{\Delta}_i \slashed{\Delta}_j  \right) = \mathcal{I}[p^2]_{ij}^{11}
+ m_i m_j \mathcal{I}[p^0]_{ij}^{11},  \\
g_2^{i(j)} + \delta  g_2^{ij} &= \frac{1}{n_D} \int [d^d p] \, {\rm tr_s} \left[ \slashed{\Delta}_i \left( i \gamma^5 \right) \slashed{\Delta}_j \left(i \gamma^5 \right) \right], \\  
g_2^{i(j)}  &= \mathcal{I}[p^2]_{ij}^{11} - m_i m_j \mathcal{I}[p^0]_{ij}^{11}, \quad \delta  g_2^{ij} = \epsilon \, \mathcal{I}[p^2]_{ij}^{11} = m_i^2 + m_j^2,
\end{split}
\end{align}
It is interesting to remark that the coefficients of the $\mathcal{O}(S^2)$ and $\mathcal{O}(P^2)$ terms are related: $g_2^{i(j)}$ is equal to $g_2^{ij}$ with the sign of $m_j$ flipped. This equivalence follows from the identity (see App.~\ref{app:master-integrals} for the definition of $\hat g_{\mu\nu}$):
\begin{equation}
\label{eq:gamma5_trick}
i \gamma^5 (\slashed{p} + m) i \gamma^5 = (\slashed{p} - m) - 2 \hat g_{\mu\nu} p^\mu \gamma^\nu,
\end{equation}
where the term proportional to $\hat g_{\mu\nu}$ is responsible for generating the finite correction  $\delta g_2^{ij}$.~\footnote{Note that using a scheme in which $\gamma^5$ (naively) anticommutes with $\gamma^{\mu}$ in $d$--dimensional space would imply that $\delta g_2^{ij} = 0$.} As detailed in App.~\ref{app:master-integrals}, the term proportional to $\hat g_{\mu\nu}$ from Eq.~\eqref{eq:gamma5_trick} stems from using the so--called BMHV scheme to handle $\gamma^5$ in $d$--dimensional space, and it needs to be kept as we are dealing with divergent loop integrals. 

Using the equality from Eq.~\eqref{eq:gamma5_trick} (or variations thereof) and trace symmetry arguments, the coefficients of terms involving even powers of $P$ can be straightforwardly expressed with the help of coefficients of terms involving only $S$, as illustrated later on. As an added bonus, for terms of dimension 5 and 6 we can set $\hat g_{\mu\nu} \to 0$, as the corresponding loop integrals are finite, which will greatly simplify our computation.

\subsection{Dimension--3 Terms} \label{subsec:dim3}

Going forward to $n=3$, Lorentz and gauge symmetries restrict the possible terms down to $\mathcal{O}(S^3, S P^2)$, which from the physical point of view renormalize the trilinear light scalar couplings present in the unbroken phase. Skipping intermediate steps, the dimension--$3$ Lagrangian is given by:
\begin{equation}
\label{eq:lagr-n=3}
16 \pi^2 \mathcal{L}_H^{n=3}= c_f \, n_D \left[ \frac{1}{3} \,  g_3^{ijk} \, {\rm tr_g} \left( S_{ij} S_{jk} S_{ki} \right) + \left( g_3^{ij(k)} + \delta g_3^{ij} \right) \, {\rm tr_g} \left( S_{ij} P_{jk} P_{ki} \right) \right],
\end{equation}
with the following coefficients:
\begin{align}
\label{eq:g3}
\begin{split}
g_3^{ijk} &= \frac{1}{n_D} \int [d^d p] \, {\rm tr_s} \left( \slashed{\Delta}_i \slashed{\Delta}_j \slashed{\Delta}_k \right) = (m_i + m_j + m_k) \, \mathcal{I}[p^2]_{ijk}^{111} + m_i m_j m_k \, \mathcal{I}[p^0]_{ijk}^{111}, \\
\delta g_3^{ij} &= (m_i + m_j) \epsilon \, \mathcal{I}[p^2]_{ijk}^{111} = m_i + m_j,
\end{split}
\end{align}
and $g_3^{ij(k)} = g_3 (m_i, m_j, - m_k)$, cf. Eq.~\eqref{eq:univ-coeff-notation-2}. As pointed out at the end of Sec.~\ref{subsec:dim2}, the coefficient of $\mathcal{O}(S P^2)$ is related to the coefficient of $\mathcal{O}(S^3)$ by flipping the sign of $m_k$ and adding a finite contribution $\delta g_3^{ij}$, which follows from of Eq.~\eqref{eq:gamma5_trick}. The symmetry factor $\frac{1}{3}$ of the gauge trace ${\rm tr_g} (S^3)$ comes from its $\mathbb{Z}_3$ symmetry, while the lack of cyclical symmetry of ${\rm tr_g} (SP^2)$ explains why it has no symmetry factor.

\subsection{Dimension--4 Terms} \label{subsec:dim4}

The discussion becomes more involved when passing to dimension--4 or higher terms, as terms with covariant derivatives are now allowed by gauge invariance, unlike the case of $n=1,2,3$. We organize the three possible terms as
\begin{equation}
\label{eq:lagr-n=4}
\mathcal{L}_H^{n=4} = \mathcal{L}_{X^4} + \mathcal{L}_{X^2 D^2} + \mathcal{L}_{D^4},
\end{equation}
where $X$ is used to generically denote $S$ and $P$. From the physical point of view, the three terms renormalize the quartic scalar couplings, the kinetic terms of the scalars, and the gauge kinetic terms, respectively.

As this is the first time we encounter traces containing covariant derivatives, we comment on whether the cyclic property can be used for such traces. When dealing with the trace over internal degrees of freedom only, denoted as $\rm tr$, the cyclic property obviously does not hold. However, this property does hold when using the full trace, $\rm Tr$, which includes a trace over the coordinate space in which the derivative operator $\partial_\mu$ acts. Since $S_H = \int d^d x \, \mathcal{L}_H$, one can convert the trace over internal degrees of freedom to a full trace~\cite{Chan:1986jq,Fuentes-Martin:2016uol} using the identity:
\begin{equation}
\label{eq:trace-to-fulltrace}
\int d^d x \, {\rm tr}[f(x)] = \frac{1}{V_d} \int d^d x \, {\rm tr}[f(x)] \, \delta^d(0) = \frac{1}{V_d} \int d^d x \, {\rm tr}\left[ \langle x | f(\hat{x}) | x \rangle \right] = \frac{1}{V_d} {\rm Tr} f(\hat{x}),
\end{equation}
where we have used the $d$--dimensional space--time volume $V_d$ to compensate the infinite Dirac distribution, $ V_d = \delta^d(0) = \langle x | x \rangle$. Using this trick, one can switch from $\rm tr$ to $\rm Tr$, apply the cyclic property for covariant derivative terms to cast them into the desired form, and then revert to $\rm tr$. The net result is that one can safely apply the cyclic property for covariant derivative terms too. This is the reason why it is possible to set up a diagrammatic approach, as done in Ref.~\cite{Zhang:2016pja}.

\vspace*{3mm}

\noindent\textbf{$\mathcal{O}(X^4)$ terms.} Expressing the $\mathcal{O}(X^4)$ terms is a straightforward generalization of the case of $\mathcal{O}(X^3)$ terms discussed in Sec.~\ref{subsec:dim3}. In the present case, however, there are more independent terms~\footnote{In this work, we define independent terms as terms which are not a cyclic permutation of one another.} involving $S$ and $P$, namely $S^4$, $S^2 P^2$, $(SP)^2$, and $P^4$. These terms are given by:
\begin{align}
\label{eq:lagr-X^4}
16 \pi^2 \mathcal{L}_{X^4} &= c_f \, n_D  \left[  \frac{1}{4} \,  g_4^{ijkl} \, {\rm tr_g} \left( S_{ij} S_{jk} S_{kl} S_{li} \right) +   \left( g_4^{ijk(l)} + \delta g_{4a} \right) \, {\rm tr_g} \left( S_{ij} S_{jk} P_{kl} P_{li} \right) \right. \notag \\ 
 & - \left.  \frac{1}{2} \,  g_4^{ij(k)(l)} \, {\rm tr_g} \left( S_{ij} P_{jk} S_{kl} P_{li} \right) + \frac{1}{4} \,  \left( g_4^{i(j)k(l)} + \delta g_{4b} \right)\, {\rm tr_g} \left( P_{ij} P_{jk} P_{kl} P_{li} \right) \right].
\end{align}
The universal coefficient $g_4^{ijkl}$ and the finite corrections $\delta g_{4a}, \delta g_{4b}$ read
\begin{align}
\label{eq:g4}
g_4^{ijkl} &= \frac{1}{n_D} \int [d^d p] \, {\rm tr_s} \left( \slashed{\Delta}_i \slashed{\Delta}_j \slashed{\Delta}_k \slashed{\Delta}_l \right) = \mathcal{I}[p^4]_{ijkl}^{1111} + m_i m_j m_k m_l \, \mathcal{I}[p^0]_{ijkl}^{1111} \notag \\ 
 &+ (m_i m_j + m_i m_k + m_i m_l + m_j m_k + m_j m_l + m_k m_l ) \mathcal{I}[p^2]_{ijkl}^{1111}, \\
\delta g_{4a} &= \epsilon \, \mathcal{I}[p^4]_{ijkl}^{1111} = 1 , \quad \delta g_{4b} = \frac{8}{3} \epsilon \, \mathcal{I}[p^4]_{ijkl}^{1111} = \frac{8}{3}, \notag
\end{align}
while the remaining universal coefficients follow from their definition from Eq.~\eqref{eq:univ-coeff-notation-2}. In Eq.~\eqref{eq:lagr-X^4} the $\frac{1}{q}$ symmetry factors have been kept in accordance with the $\mathbb{Z}_q$ symmetry of each trace, while the minus sign in front of the third term is a result of the identity 
\begin{equation}
\label{eq:gamma5_trick-2}
i\gamma^5 (\slashed{p} + m_1) (\slashed{p} + m_2) i \gamma^5 =  - (\slashed{p} - m_1)(\slashed{p} - m_2) - 2 (m_1 + m_2) \hat{g}_{\mu\nu} p^\mu \gamma^\nu
\end{equation}
which is a simple variation of Eq.~\eqref{eq:gamma5_trick}. Note that the $g_4^{ij(k)(l)}$ term does not receive any finite corrections stemming from the BMHV treatment of $\gamma^5$, as the term proportional to $\hat{g}_{\mu\nu}$ from Eq.~\eqref{eq:gamma5_trick-2} scales as $p^\mu$ and not $p^2$. Consequently, $\hat{g}_{\mu\nu}$ ends up multiplying a finite integral, and the result vanishes when taking the limit $\epsilon \to 0$. 

\vspace*{3mm}

\noindent\textbf{$\mathcal{O}(X^2 D^2)$ terms.} For calculating the $\mathcal{O}(X^2 D^2)$ terms, we use an approach similar to the one presented in Ref.~\cite{Zhang:2016pja}. We focus on the $\mathcal{O}(S^2 D^2)$ term, from which the remaining $\mathcal{O}(P^2 D^2)$ term follows immediately, as mentioned at the end of Sec.~\ref{subsec:dim2}.

We first note that there are two independent terms that contain two covariant derivatives and two powers of $S$, $S^2 D^2$ and $(SD)^2$, and their coefficients follow from the $n=4$ term in the Taylor expansion of Eq.~\eqref{eq:effective-lagrangian}:
\begin{align}
\label{eq:lagr-D^2_S^2-aux}
16 \pi^2 \mathcal{L}_{S^2 D^2} &= - c_f \left[ \int [d^d p] \, {\rm tr_s} \left( \slashed{\Delta}_i \slashed{\Delta}_j \slashed{\Delta}_i \gamma_\mu \slashed{\Delta}_i \gamma_\nu \right) {\rm tr_g} \left( S_{ij} S_{ji} D^\mu_i D^\nu_i \right) \right. \notag \\
&+ \left. \frac{1}{2} \int [d^d p] \, {\rm tr_s} \left( \slashed{\Delta}_i \slashed{\Delta}_j \gamma_\mu \slashed{\Delta}_j  \slashed{\Delta}_i \gamma_\nu \right) {\rm tr_g} \left( S_{ij} D^\mu_j S_{ji}  D^\nu_i \right) \right].
\end{align}
Note that each term has the appropriate symmetry factor.  Afterwards, we write down all the possible gauge--invariant terms (and the corresponding symmetry factors) arising at $\mathcal{O}(S^2 D^2)$ -- in this case there is only term -- and expand the commutators:
\begin{align}
\label{eq:lagr-D^2_S^2-aux2}
16 \pi^2 \mathcal{L}_{S^2 D^2} &= \frac{1}{2} c_f \, n_D \,  g_5^{ij} \, {\rm tr_g} \left( \left[ D_\mu , S \right]_{ij} \left[ D^\mu , S \right]_{ji}  \right) \notag \\
&= c_f \, n_D \,  g_5^{ij} {\rm tr_g} \left(  S_{ij} D^{\mu}_j S_{ji}  D_{\mu,i} - S_{ij} S_{ji} D^2_i \right),
\end{align}
where we have used the symmetry of the form factor, $g_5^{ij} = g_5^{ji}$, which follows from the symmetry of the associated trace. With Eqs.~\eqref{eq:lagr-D^2_S^2-aux} and \eqref{eq:lagr-D^2_S^2-aux2} at our disposal, we can solve for $g_5^{ij}$ by equating the factors multiplying the $(SD)^2$ term:
\begin{align}
\label{eq:g5}
g_5^{ij} &= -\frac{1}{2 \, n_D} \frac{g^{\mu\nu}}{d} \int [d^d p] \, {\rm tr_s} \left( \slashed{\Delta}_i \slashed{\Delta}_j \gamma_\mu \slashed{\Delta}_j  \slashed{\Delta}_i \gamma_\nu \right) \notag \\
 &= - \frac{1}{2} \mathcal{I}[p^4]_{ij}^{22} + \frac{(m_i - m_j)^2}{4} \mathcal{I}[p^2]_{ij}^{22} - \frac{m_i^2 m_j^2}{2} \mathcal{I}[p^0]_{ij}^{22},
\end{align}
with $d = 4 - 2 \epsilon$ the number of space--time dimensions. We mention that, equivalently, $g_5^{ij}$ could have been computed by matching the factors in front of the $S^2 D^2$ term. While for calculating $g_5^{ij}$ it is sufficient to equate the prefactors of only one of the two independent terms~\footnote{This redundancy is also pointed out in Ref.~\cite{Zhang:2016pja} for the case of covariant diagrams, and has the benefit of drastically reducing the number of diagrams that need to be computed.} in Eqs.~\eqref{eq:lagr-D^2_S^2-aux} and \eqref{eq:lagr-D^2_S^2-aux2}, considering the other term does have some value, as it provides a consistency check of the results. We have performed this check and re--confirmed Eq.~\eqref{eq:g5}. 

As for the $\mathcal{O}(P^2 D^2)$ term, the computation proceeds in an equivalent manner, which allows us to write the effective Lagrangian at $\mathcal{O}(X^2 D^2 )$:
\begin{equation}
\label{eq:lagr-D^2_X^2}
16 \pi^2 \mathcal{L}_{X^2 D^2} = \frac{1}{2} c_f \, n_D \left[  g_5^{ij} \, {\rm tr_g} \left( \left[ D_\mu , S \right]_{ij} \left[ D^\mu , S \right]_{ji} \right) + \left( g_5^{i(j)} + \delta g_5 \right) \, {\rm tr_g} \left( \left[ D_\mu , P \right]_{ij} \left[ D^\mu , P \right]_{ji}  \right) \right],
\end{equation}
where the finite correction is given by
\begin{equation}
\label{eq:delta-g5}
\delta g_5 = - \frac{\epsilon}{2} \mathcal{I}[p^4]_{ij}^{22} = -\frac{1}{2}.
\end{equation}

\vspace*{3mm}

\noindent\textbf{$\mathcal{O}(D^4)$ terms.} Since the $D_\mu$'s are diagonal in multiplet space, the term involving four covariant derivatives depends only on one mass. The only gauge invariant quantity involving four $D_\mu$'s is $F^{\mu\nu} F_{\mu\nu} $, with $F_{\mu\nu}$ defined in Eq.~\eqref{eq:F-munu}, hence the $\mathcal{O}(D^4)$ term reads
\begin{equation}
\label{eq:lagr-D^4}
16 \pi^2 \mathcal{L}_{D^4} =  \frac{1}{2}  c_f \, n_D \, g_6^i \, {\rm tr_g} \left( F_i^{\mu\nu} F_{\mu\nu,i} \right),
\end{equation}
with
\begin{align}
\label{eq:g6}
g_6^i &= \frac{g^{\mu\nu} g^{\rho\sigma}}{2 \, n_D \, d(d-1)} \int [d^d p] \, {\rm tr_s} \left( \slashed{\Delta}_i \gamma_\mu \slashed{\Delta}_i \gamma_\nu \slashed{\Delta}_i \gamma_\rho \slashed{\Delta}_i \gamma_\sigma \right) \notag \\
 &= \frac{1}{6}\left(1 + \frac{5  \epsilon}{6} \right) \mathcal{I}[p^4]_i^4 - \frac{1}{2} m_i^2 \mathcal{I}[p^2]_i^4 + \frac{2}{3} m_i^4 \mathcal{I}[p^0]_i^4,
\end{align}
where the $\mathcal{O}(\epsilon)$ term in the factor multiplying the divergent integral $\mathcal{I}[p^4]_i^4$ was retained to correctly account for the finite part. We note that our result for the $\mathcal{O}(D^4)$ term agrees with the findings of Refs.~\cite{Henning:2014wua,Zhang:2016pja}.

For clarity, let us break down the intermediate steps in obtaining the result in Eq.~\eqref{eq:lagr-D^4}. Starting from the Taylor expansion in Eq.~\eqref{eq:effective-lagrangian}, the $\mathcal{O}(D^4)$ piece is given by:
\begin{align}
\label{eq:lagr-D^4-details}
16 \pi^2 \mathcal{L}_{D^4} &= \frac{1}{4} c_f \int [d^d p] \, {\rm tr_s} \left( \slashed{\Delta}_i \gamma_\mu \slashed{\Delta}_i \gamma_\nu \slashed{\Delta}_i \gamma_\rho \slashed{\Delta}_i \gamma_\sigma \right) {\rm tr_g} (D^\mu_i D^\nu_i D^\rho_i D^\sigma_i) \notag \\
&=  \frac{1}{2}  c_f \, n_D \, g_6^i  \left( g_{\mu\nu} g_{\rho\sigma} -2 g_{\mu\rho} g_{\nu\sigma} + g_{\mu\sigma} g_{\nu\rho} \right) {\rm tr_g} (D^\mu_i D^\nu_i D^\rho_i D^\sigma_i) \notag \\
&=  \frac{1}{2}  c_f \, n_D \, g_6^i \,  {\rm tr_g} (2 D^\mu_i D^2_i D_{\mu,i} - 2 D^\mu_i D^\nu_i D_{\mu,i} D_{\nu,i} ).
\end{align}
Passing to the third line is done through a cyclic permutation of the $D^2_i D^2_i$ term, and then  Eq.~\eqref{eq:lagr-D^4} is recovered by noting that $2 D^\mu_i D^\nu_i D_{\nu,i} D_{\mu,i} - 2 D^\mu_i D^\nu_i D_{\mu,i} D_{\nu,i} =  F^{\mu\nu,i} F_{\mu\nu}^i$.


\subsection{Dimension--5 Terms} \label{subsec:dim5}

For the dimension--5 case, we organize the possible terms as follows:
\begin{equation}
\label{eq:lagr-n=5}
\mathcal{L}_H^{n=5} = \mathcal{L}_{X^5} + \mathcal{L}_{X^3 D^2} + \mathcal{L}_{S D^4} + \mathcal{L}_{P D^4}.
\end{equation}
We choose to treat the $\mathcal{O}(S D^4)$ and $\mathcal{O}(P D^4)$ terms separately, as they are different in terms of their $CP$ properties. 

As pointed out previously, the dimension--5 and dimension--6 universal coefficients are finite, allowing us to compute the corresponding loop integrals in 4 instead of $d$ dimensions. Therefore, $\gamma^5$ retains its usual anticommuting properties and we no longer need to keep the pieces proportional to $\hat{g}_{\mu\nu}$ from Eqs.~(\ref{eq:gamma5_trick}, \ref{eq:gamma5_trick-2}) (or variations thereof) when computing the universal coefficients for terms involving even powers of $P$. We stress once again that, with the help of Eqs.~(\ref{eq:gamma5_trick}, \ref{eq:gamma5_trick-2}) and trace symmetry arguments, these coefficients follow effortlessly from the coefficients of operators containing only insertions of $S$.

\vspace*{3mm}

\noindent\textbf{$\mathcal{O}(X^5)$ terms.} Similarly to the $\mathcal{O}(X^3)$ and $\mathcal{O}(X^4)$ contributions, the $\mathcal{O}(X^5)$ Lagrangian reads:
\begin{align}
\label{eq:lagr-X^5}
16 \pi^2 \mathcal{L}_{X^5} = c_f \, n_D & \left[  \frac{1}{5} \,  g_7^{ijklm} \, {\rm tr_g} \left( S_{ij} S_{jk} S_{kl} S_{lm} S_{mi} \right) +   g_7^{ijkl(m)} \, {\rm tr_g} \left( S_{ij} S_{jk} S_{kl} P_{lm} P_{mi} \right) \right. \notag \\ 
 & - \left.   g_7^{ijk(lm)} \, {\rm tr_g} \left( S_{ij} S_{jk} P_{kl} S_{lm} P_{mi} \right) + g_7^{ij(k)l(m)} \, {\rm tr_g} \left( S_{ij} P_{jk} P_{kl} P_{lm} P_{mi} \right) \right],
\end{align}
with the universal coefficient given by:
\begin{align}
\label{eq:g7}
g_7^{ijklm} &= \frac{1}{n_D} \int [d^d p] \, {\rm tr_s} \left( \slashed{\Delta}_i \slashed{\Delta}_j \slashed{\Delta}_k \slashed{\Delta}_l \slashed{\Delta}_m \right) \notag \\
&= (m_i + m_j + m_k + m_l + m_m) \, \mathcal{I}[p^4]_{ijklm}^{11111} + [ m_i m_j ( m_k + m_l + m_m )   \notag \\ 
 &+ (m_i + m_j) m_k ( m_l + m_m ) + (m_i + m_j + m_k) m_l m_m  ] \, \mathcal{I}[p^2]_{ijklm}^{11111} \notag \\
 & + m_i m_j m_k m_l  m_m \, \mathcal{I}[p^0]_{ijklm}^{11111} .
\end{align}

\vspace*{3mm}

\noindent\textbf{$\mathcal{O}(X^3 D^2)$ terms.} For the $\mathcal{O}(X^3 D^2)$ terms, we follow the same procedure as for the case of $\mathcal{O}(X^2 D^2)$ terms. Focusing on the  $\mathcal{O}(S^3 D^2)$ contribution, the only independent gauge--invariant combination is
\begin{align}
\label{eq:lagr-D^2_S^3-aux}
16 \pi^2 \mathcal{L}_{S^3 D^2} &= c_f \, n_D \, g_8^{ijk} {\rm tr_g} \left( S_{ij} \left[ D_\mu, S \right]_{jk} \left[ D^\mu, S \right]_{ki} \right) \notag \\
&= c_f \, n_D \, {\rm tr_g} \left[ \left( g_8^{ijk} + g_8^{jki} - g_8^{kij} \right) \left( S_{ij} S_{jk} D^\mu_k S_{ki} D_{\mu,i} \right) - g_8^{ijk} \left( S_{ij} S_{jk} D^2_k S_{ki}  \right) \right].
\end{align}
It is clear from this relation that the easiest way to find $g_8^{ijk}$ is to compute the loop integral multiplying ${\rm tr_g} \left(  S_{ij} S_{jk} D^2_k S_{ki} \right)$ from the covariant derivative expansion in Eq.~\eqref{eq:effective-lagrangian}:
\begin{equation}
\label{eq:lagr-D^2_S^3-aux2}
16 \pi^2 \mathcal{L}_{S^3 D^2} \supset -c_f \int [d^d p] \, {\rm tr_s} \left( \slashed{\Delta}_i \slashed{\Delta}_j \slashed{\Delta}_k \gamma_\mu \slashed{\Delta}_k \gamma_\nu \slashed{\Delta}_k \right) {\rm tr_g} \left( S_{ij} S_{jk} D^\mu_k D^\nu_k S_{ki} \right),
\end{equation}
from which $g_8^{ijk}$ is found to be~\footnote{Since for dimension--5 and higher the loop integrals are always finite, we are dividing by $4$ instead of $d=4-2\epsilon$.}
\begin{align}
\label{eq:g8}
g_8^{ijk} &= \frac{1}{n_D} \frac{g^{\mu\nu}}{4} \int [d^d p] \, {\rm tr_s} \left( \slashed{\Delta}_i \slashed{\Delta}_j \slashed{\Delta}_k \gamma_\mu \slashed{\Delta}_k \gamma_\nu \slashed{\Delta}_k \right) \notag \\
 &= - \frac{m_i + m_j}{2} \mathcal{I}[p^4]_{ijk}^{113} + \frac{(3 m_i + 3 m_j + 2 m_k) m_k^2}{2} \mathcal{I}[p^2]_{ijk}^{113} + m_i m_j m_k^3 \, \mathcal{I}[p^0]_{ijk}^{113}.
\end{align}
Including the terms containing powers of $P$, the $\mathcal{O}(X^3 D^2)$ Lagrangian reads:
\begin{align}
\label{eq:lagr-D^2_S^3}
16 \pi^2 \mathcal{L}_{X^3 D^2} &= c_f \, n_D \left[ g_8^{ijk} {\rm tr_g} \left( S_{ij} \left[ D_\mu, S \right]_{jk} \left[ D^\mu, S \right]_{ki} \right) + g_8^{ij(k)} {\rm tr_g} \left( S_{ij} \left[ D_\mu, P \right]_{jk} \left[ D^\mu, P \right]_{ki} \right) \right. \notag \\
& \left. + g_8^{(i)jk} {\rm tr_g} \left( P_{ij} \left[ D_\mu, S \right]_{jk} \left[ D^\mu, P \right]_{ki} \right) + g_8^{i(j)k} {\rm tr_g} \left( P_{ij} \left[ D_\mu, P \right]_{jk} \left[ D^\mu, S \right]_{ki} \right) \right].
\end{align}
\vspace*{3mm}

\noindent\textbf{$\mathcal{O}(S D^4)$ terms.} The simplest way to compute this is to directly compute it from the CDE in Eq.~\eqref{eq:effective-lagrangian}. The result is:
\begin{align}
\label{eq:lagr-D^4_S}
16 \pi^2 \mathcal{L}_{S D^4} &= c_f \int [d^d p] \, {\rm tr_s} \left( \slashed{\Delta}_i \slashed{\Delta}_i \gamma_\mu  \slashed{\Delta}_i \gamma_\nu  \slashed{\Delta}_i \gamma_\rho \slashed{\Delta}_i \gamma_\sigma \right) {\rm tr_g} \left( S_{ii} D^{\mu}_i D^{\nu}_i D^{\rho}_i D^{\sigma}_i \right) \notag \\
&= - c_f \frac{4}{3 m_i} \left( g_{\mu\sigma} g_{\nu\rho} - g_{\mu\rho} g_{\nu\sigma}  \right) {\rm tr_g} \left( S_{ii} D^{\mu}_i D^{\nu}_i D^{\rho}_i D^{\sigma}_i \right) \notag \\
& = c_f n_D \left( - \frac{1}{6 m_i} \right) {\rm tr_g} \left( S_{ii} F^{\mu\nu}_i F_{\mu\nu,i} \right),
\end{align}
from where the universal coefficient $g_9^i$ can be easily read off as:
\begin{equation}
\label{eq:g9}
g_9^i = - \frac{1}{6 m_i}.
\end{equation}
We chose to write $g_9^i$ explicitly, as expressing it through the master integrals defined in Eq.~\eqref{eq:master-integrals} would have lead to a much more cumbersome relation.

Note that the $\mathcal{O}(SD^4)$ term from Eq.~\eqref{eq:lagr-D^4_S} depends only on the diagonal entries of $S$. From the physical point of view, it means that this dimension--5 term is generated only if the theory contains light real (pseudo)scalars. In the case of pseudoscalars, the term from Eq.~\eqref{eq:lagr-D^4_S} would violate $CP$--symmetry.

\vspace*{3mm}

\noindent\textbf{$\mathcal{O}(P D^4)$ terms.} Similarly to $\mathcal{O}(SD^4)$, the $\mathcal{O}(P D^4)$ term can be easily computed directly from Eq.~\eqref{eq:effective-lagrangian}:
\begin{align}
\label{eq:lagr-D^4_P}
16 \pi^2 \mathcal{L}_{P D^4} &= i \, c_f \int [d^d p] \, {\rm tr_s} \left( \slashed{\Delta}_i \gamma^5 \slashed{\Delta}_i \gamma_\mu  \slashed{\Delta}_i \gamma_\nu  \slashed{\Delta}_i \gamma_\rho \slashed{\Delta}_i \gamma_\sigma \right) {\rm tr_g} \left( P_{ii} D^{\mu}_i D^{\nu}_i D^{\rho}_i D^{\sigma}_i \right) \notag \\
&= -  c_f \frac{2}{m_i} \varepsilon_{\mu\nu\rho\sigma} {\rm tr_g} \left( P_{ii} D^{\mu}_i D^{\nu}_i D^{\rho}_i D^{\sigma}_i \right) = c_f n_D \left( \frac{1}{4 m_i} \right) {\rm tr_g} \left( P_{ii} \widetilde{F}^{\mu\nu}_i F_{\mu\nu,i} \right),
\end{align}
where $\varepsilon_{\mu\nu\rho\sigma}$ is the 4--dimensional Levi--Civita tensor, and we have defined
\begin{equation}
\label{eq:F-tilde-munu}
\widetilde{F}_{\mu\nu} \equiv \frac{1}{2} \varepsilon_{\mu\nu\rho\sigma} F^{\rho\sigma}.
\end{equation}
The universal coefficient appearing in Eq.~\eqref{eq:lagr-D^4_P} is:
\begin{equation}
\label{eq:g10}
g_{10}^i = - \frac{m_i}{2} \mathcal{I}[p^0]_i^3 = \frac{1}{4 m_i},
\end{equation}
which we also write down in terms of master integrals thanks to the compactness of the expression.
 
As in the $\mathcal{O}(SD^4)$ case, the $\mathcal{O}(P D^4)$ term vanishes unless there are real (pseudo)scalars present in the light particle spectrum. Contrary to $\mathcal{O}(SD^4)$, the term in Eq.~\eqref{eq:lagr-D^4_P} conserves $CP$ in the case of light pseudoscalars.

\subsection{Dimension--6 Terms} \label{subsec:dim6}

We now turn our attention towards the final set of terms considered in this work, the dimension--6 terms. Gauge invariance allows for a multitude of $n=6$ terms, which we organize as:
\begin{equation}
\label{eq:lagr-n=6}
\mathcal{L}_H^{n=6} = \mathcal{L}_{X^6} + \mathcal{L}_{X^4 D^2} + \mathcal{L}_{X^2 D^4} + \mathcal{L}_{SP D^4} + \mathcal{L}_{D^6}.
\end{equation}
The $\mathcal{O}(SP D^4)$ piece is written separately from $\mathcal{O}(X^2 D^4)$ because it is the only one at this dimension that depends on the dual field strength tensor $\widetilde{F}_{\mu\nu}$.

\vspace*{3mm}

\noindent\textbf{$\mathcal{O}(X^6)$ terms.} Although lengthy, the $X^6$ piece is straightforward to compute by generalizing from $\mathcal{O}(X^5)$ and reads: 
\begin{align}
\label{eq:lagr-X^6}
16 \pi^2 \mathcal{L}_{X^6} &= c_f \, n_D  \left[  \frac{1}{6} \,  g_{11}^{ijklmn} \, {\rm tr_g} \left( S_{ij} S_{jk} S_{kl} S_{lm} S_{mn} S_{ni} \right) + g_{11}^{ijklm(n)} \, {\rm tr_g} \left( S_{ij} S_{jk} S_{kl} S_{lm} P_{mn} P_{ni} \right) \right. \notag \\
& \left. -  g_{11}^{ijkl(m)(n)} \, {\rm tr_g} \left( S_{ij} S_{jk} S_{kl} P_{lm} S_{mn} P_{ni} \right)   + \frac{1}{2} \,  g_{11}^{ijk(l)(m)(n)} \, {\rm tr_g} \left( S_{ij} S_{jk} P_{kl} S_{lm} S_{mn} P_{ni} \right) \right. \notag \\
& \left.  + g_{11}^{ijk(l)m(n)} \, {\rm tr_g} \left( S_{ij} S_{jk} P_{kl} P_{lm} P_{mn} P_{ni} \right) - g_{11}^{ij(k)(l)m(n)} \, {\rm tr_g} \left( S_{ij} P_{jk} S_{kl} P_{lm} P_{mn} P_{ni} \right) \right. \notag \\
& \left. + \frac{1}{2} \,  g_{11}^{ij(k)lm(n)} \, {\rm tr_g} \left( S_{ij} P_{jk} P_{kl} S_{lm} P_{mn} P_{ni} \right) + \frac{1}{6} \,  g_{11}^{i(j)k(l)m(n)} \, {\rm tr_g} \left( P_{ij} P_{jk} P_{kl} P_{lm} P_{mn} P_{ni} \right) \right],
\end{align}
while the universal coefficient is equal to:
\begin{align}
\label{eq:g11}
g_{11}^{ijklmn} &= \frac{1}{n_D} \int [d^d p] \, {\rm tr_s} \left( \slashed{\Delta}_i \slashed{\Delta}_j \slashed{\Delta}_k \slashed{\Delta}_l \slashed{\Delta}_m \slashed{\Delta}_n \right) \notag \\
& = \mathcal{I}[p^6]_{ijklmn}^{111111} + a_+  \mathcal{I}[p^4]_{ijklmn}^{111111} +  m_i m_j m_k m_l m_m m_n \left(  a_- \mathcal{I}[p^2]_{ijklmn}^{111111} + \mathcal{I}[p^0]_{ijklmn}^{111111} \right).
\end{align}
To maintain the expression compact, we have defined $a_{\pm}$ as:
\begin{align*}
a_{\pm} &= (m_i m_j)^{\pm 1} + (m_i m_k)^{\pm 1}  + (m_i m_l)^{\pm 1} + (m_i m_m)^{\pm 1} + (m_i m_n)^{\pm 1} \\ 
&+ (m_j m_k)^{\pm 1} + (m_j m_l)^{\pm 1} + (m_j m_m)^{\pm 1} + (m_j m_n)^{\pm 1} + (m_k m_l)^{\pm 1} \\ 
& + (m_k m_m)^{\pm 1} + (m_k m_n)^{\pm 1} + (m_l m_m)^{\pm 1} + (m_l m_n)^{\pm 1} + (m_m m_n)^{\pm 1}.
\end{align*}

The $\mathcal{O}(X^6)$ terms suggestively illustrate how the use of trace symmetry and of Eqs.~\eqref{eq:gamma5_trick} and \eqref{eq:gamma5_trick-2} streamlines our computation. Instead of having to calculate eight operators and their corresponding universal coefficients, it is enough to consider only one operator, $ S^6 $, together its universal coefficient, with the other seven following effortlessly. 
 
\vspace*{3mm}

\noindent\textbf{$\mathcal{O}(X^4 D^2)$ terms.} As in the case of the other terms with two covariant derivatives, we first  compute the universal coefficients for $P=0$, and then generalize to $P \neq 0$. At $\mathcal{O}(S^4 D^2)$, we have two gauge invariant terms, which upon expanding the commutators become:
\begin{align}
\label{eq:lagr-D^2_S^4-aux}
16 \pi^2 \mathcal{L}_{S^4 D^2} &= c_f \, n_D \left[ g_{12}^{ijkl} {\rm tr_g} \left( S_{ij} S_{jk} \left[ D_\mu, S \right]_{kl} \left[ D^\mu, S \right]_{li} \right) + \frac{1}{2} g_{13}^{ijkl} {\rm tr_g} \left( S_{ij} \left[ D_\mu, S \right]_{jk} S_{kl} \left[ D^\mu, S \right]_{li} \right)  \right] \notag \\
&= c_f \, n_D \, {\rm tr_g} \left[ - g_{12}^{ijkl} \left( S_{ij} S_{jk} S_{kl} D^2_l S_{li}  \right) + \left( g_{12}^{ijkl} + g_{12}^{jkli} - g_{13}^{jkli} \right) \left( S_{ij} S_{jk} S_{kl} D^\mu_l S_{li} D_{\mu,i} \right) \right. \notag \\
&\left. \quad + \frac{1}{2} \left( g_{13}^{ijkl} + g_{13}^{jkli} - g_{12}^{ijkl} - g_{12}^{klij}  \right) \left( S_{ij} S_{jk} D^\mu_k S_{kl}  S_{li} D_{\mu,i} \right)  \right].
\end{align}
In the last equality, we have used the relation $g_{13}^{ijkl} = g_{13}^{klij}$ (inherited from the symmetry of the associated trace) and performed the symmetrization:
\begin{equation*}
g_{12}^{ijkl}  \left( S_{ij} S_{jk} D^\mu_k S_{kl}  S_{li} D_{\mu,i} \right) = \frac{1}{2} \left( g_{12}^{ijkl} + g_{12}^{klij}  \right) \left( S_{ij} S_{jk} D^\mu_k S_{kl}  S_{li} D_{\mu,i} \right).
\end{equation*}
To calculate $g_{12,13}^{ijkl}$, we match Eq.~\eqref{eq:lagr-D^2_S^4-aux} on the corresponding terms obtained form the CDE in Eq.~\eqref{eq:effective-lagrangian}, which we choose to be
\begin{align}
\label{eq:lagr-D^2_S^4-aux2}
16 \pi^2 \mathcal{L}_{S^4 D^2}  \supset -c_f &\left[ \int [d^d p] \, {\rm tr_s} \left( \slashed{\Delta}_i \slashed{\Delta}_j \slashed{\Delta}_k \slashed{\Delta}_l \gamma_\mu \slashed{\Delta}_l \gamma_\nu \slashed{\Delta}_l \right) {\rm tr_g} \left( S_{ij} S_{jk} S_{kl} D^\mu_l D^\nu_l S_{li} \right), \right.  \notag \\
& \left. + \int [d^d p] \, {\rm tr_s} \left( \slashed{\Delta}_i \slashed{\Delta}_j \slashed{\Delta}_k \slashed{\Delta}_l \gamma_\mu \slashed{\Delta}_l \slashed{\Delta}_i \gamma_\nu \right) {\rm tr_g} \left( S_{ij} S_{jk} S_{kl} D^\mu_l S_{li} D^\nu_i \right) \right].
\end{align}
Equating the expressions in Eqs.~\eqref{eq:lagr-D^2_S^4-aux} and \eqref{eq:lagr-D^2_S^4-aux2}, we find:
\begin{align}
\label{eq:g12}
g_{12}^{ijkl} &= \frac{1}{n_D} \frac{g^{\mu\nu}}{4}   \int [d^d p] \, {\rm tr_s} \left( \slashed{\Delta}_i \slashed{\Delta}_j \slashed{\Delta}_k \slashed{\Delta}_l \gamma_\mu \slashed{\Delta}_l \gamma_\nu \slashed{\Delta}_l \right) \notag \\
&= -\frac{1}{2} \mathcal{I}[p^6]_{ijkl}^{1113} - \frac{1}{2} (m_i m_j + m_i m_k + m_j m_k - 3 m_l^2) \mathcal{I}[p^4]_{ijkl}^{1113} \notag \\
&+ \left[ \frac{3}{2} (m_i m_j + m_i m_k + m_j m_k) + (m_i + m_j + m_k) m_l \right] m_l^2 \, \mathcal{I}[p^2]_{ijkl}^{1113} \notag \\ 
&+ m_i m_j m_k m_l^3 \, \mathcal{I}[p^0]_{ijkl}^{1113},
\end{align}
and
\begin{align}
\label{eq:g13}
g_{13}^{ijkl} &= g_{12}^{ijkl} + g_{12}^{lijk} + \frac{1}{n_D} \frac{g^{\mu\nu}}{4}   \int [d^d p] \, {\rm tr_s} \left( \slashed{\Delta}_i \slashed{\Delta}_j \slashed{\Delta}_k \gamma_\mu \slashed{\Delta}_k  \slashed{\Delta}_l \gamma_\nu \slashed{\Delta}_l \right) \notag \\
&= g_{12}^{ijkl} + g_{12}^{lijk} + \mathcal{I}[p^6]_{ijkl}^{1122} + \frac{1}{2} \left[ 2 m_i m_j + (m_i + m_j)(m_k + m_l) - (m_k - m_l)^2 \right] \mathcal{I}[p^4]_{ijkl}^{1122} \notag \\
&+ \frac{1}{2} \left[ - m_i m_j (m_k - m_l)^2 + (m_i + m_j)(m_k + m_l) m_k m_l + 2 m_k^2 m_l^2  \right] \mathcal{I}[p^2]_{ijkl}^{1122} \notag \\
&+ m_i m_j m_k^2 m_l^2 \, \mathcal{I}[p^0]_{ijkl}^{1122}.
\end{align}
As a check, we have also performed the matching on the remaining term, $S^2 D_\mu S^2 D_\mu$, and found it to be consistent with our expressions for $g_{12}^{ijkl}$ and $g_{13}^{ijkl}$.

Having all the ingredients, we finally write down the full $\mathcal{O}(X^4 D^2)$ effective Lagrangian, including terms with $P$ as well:
\begin{align}
\label{eq:lagr-X^4_D^2}
16 \pi^2 \mathcal{L}_{X^4 D^2} &= c_f \, n_D \bigg \lbrace {\rm tr_g} \left[ g_{12}^{ijkl}  \left( S_{ij} S_{jk} \left[ D_\mu, S \right]_{kl} \left[ D^\mu, S \right]_{li} \right) + g_{12}^{ijk(l)}  \left( S_{ij} S_{jk} \left[ D_\mu, P \right]_{kl} \left[ D^\mu, P \right]_{li} \right) \right. \notag \\ 
& \left. + g_{12}^{ij(k)l}  \left( S_{ij} P_{jk} \left[ D_\mu, P \right]_{kl} \left[ D^\mu, S \right]_{li} \right) + g_{12}^{i(j)kl}  \left( P_{ij} P_{jk} \left[ D_\mu, S \right]_{kl} \left[ D^\mu, S \right]_{li} \right) \right. \notag \\ 
& \left. + g_{12}^{(i)jkl}  \left( P_{ij} S_{jk} \left[ D_\mu, S \right]_{kl} \left[ D^\mu, P \right]_{li} \right) - g_{12}^{ij(k)(l)}  \left( S_{ij} P_{jk} \left[ D_\mu, S \right]_{kl} \left[ D^\mu, P \right]_{li} \right) \right. \notag \\ 
& \left. - g_{12}^{i(j)(k)l}  \left( P_{ij} S_{jk} \left[ D_\mu, P \right]_{kl} \left[ D^\mu, S \right]_{li} \right) + g_{12}^{i(j)k(l)}  \left( P_{ij} P_{jk} \left[ D_\mu, P \right]_{kl} \left[ D^\mu, P \right]_{li} \right)  \right] \notag \\
&+ {\rm tr_g} \left[ \frac{1}{2} g_{13}^{ijkl}  \left( S_{ij} \left[ D_\mu, S \right]_{jk} S_{kl}  \left[ D^\mu, S \right]_{li} \right) + g_{13}^{ijk(l)}  \left( S_{ij} \left[ D_\mu, S \right]_{jk} P_{kl}  \left[ D^\mu, P \right]_{li} \right) \right. \notag \\ 
& \left. +  g_{13}^{ij(k)l}  \left( S_{ij} \left[ D_\mu, P \right]_{jk} P_{kl}  \left[ D^\mu, S \right]_{li} \right) - \frac{1}{2} g_{13}^{ij(k)(l)}  \left( S_{ij} \left[ D_\mu, P \right]_{jk} S_{kl}  \left[ D^\mu, P \right]_{li} \right) \right. \notag \\ 
& \left. - \frac{1}{2} g_{13}^{i(j)(k)l}  \left( P_{ij} \left[ D_\mu, S \right]_{jk} P_{kl}  \left[ D^\mu, S \right]_{li} \right) + \frac{1}{2} g_{13}^{i(j)k(l)}  \left( P_{ij} \left[ D_\mu, P \right]_{jk} P_{kl}  \left[ D^\mu, P \right]_{li} \right)  \right] \bigg \rbrace.
\end{align}

Again, trace symmetry arguments together with Eqs.~(\ref{eq:gamma5_trick}, \ref{eq:gamma5_trick-2}) made our task much simpler: instead of fourteen different terms, we only had to compute two, namely $S^2 \left[ D_\mu, S \right]^2$ and $\left(S \left[ D_\mu, S \right] \right)^2 $.

\vspace*{3mm}

\noindent\textbf{$\mathcal{O}(X^2 D^4)$ terms.} Focusing again on the terms containing only $S$, there are four independent gauge invariant traces that arise at $\mathcal{O}(S^2 D^4)$. We choose them as follows:
\begin{align}
\label{eq:lagr-D^4_S^2-aux}
16 \pi^2 \mathcal{L}_{S^2 D^4} &= c_f \, n_D \, {\rm tr_g} \left[ \frac{1}{2} g_{14}^{ij} \left( \left[ D_\mu, [D^\mu, S] \right]_{ij} \left[ D_\nu, [D^\nu, S] \right]_{ji} \right) + g_{15}^{ij}  \left( S_{ij} S_{ji} F^{\mu\nu}_i F_{\mu\nu,i} \right) \right. \notag \\ 
& \left. + \frac{1}{2} g_{16}^{ij}  \left( S_{ij} F^{\mu\nu}_j S_{ji}  F_{\mu\nu,i} \right) + i \, g_{17}^{ij} \left( S_{ij} [D_\mu, S]_{ji}  \left[ D_\nu, F^{\nu\mu} \right]_i \right) \right] \notag \\
& \supset c_f \, n_D \, {\rm tr_g} \left[ \left( 2 g_{15}^{ij} - g_{17}^{ij} \right) \left( S_{ij} S_{ji} D^{\mu}_i D^2_i D_{\mu,i} \right) - 2 g_{17}^{ij} \left( S_{ij} D^{\mu}_j S_{ji}  D^{\nu}_i D_{\mu,i} D_{\nu,i} \right) \right. \notag \\
& \left. + g_{14}^{ij} \left( S_{ij} D^2_j S_{ji} D^2_i \right) - g_{16}^{ij} \left( S_{ij} D^{\mu}_j D^{\nu}_j S_{ji}   D_{\mu,i} D_{\nu,i} \right)  \right],
\end{align}
where we have expanded the commutators and kept only four open covariant derivative terms, which are necessary for computing the four universal coefficients present at this order. 

Before matching Eq.~\eqref{eq:lagr-D^4_S^2-aux} to the corresponding terms from the CDE in Eq.~\eqref{eq:effective-lagrangian}, we write down some useful relations for calculating the universal coefficients. At $\mathcal{O}(S^2 D^4)$ (and $\mathcal{O}(P^2 D^4)$), the loop integrals have four Lorentz indices and thus have the general form~\footnote{The 4D Levi--Civita tensor $\varepsilon_{\mu\nu\rho\sigma}$ does not appear as the spin traces involved contain an even number of $\gamma^5$ matrices.}
\begin{equation*}
a_1 \, g_{\mu\nu} g_{\rho\sigma} + a_2 \, g_{\mu\sigma} g_{\nu\rho} + a_3 \, g_{\mu\rho} g_{\nu\sigma}.
\end{equation*}
In practice, however, we do not need the full loop integral, but just its scalar components, defined above as $a_{1,2,3}$. To isolate these components, we define
\begin{equation*}
\mathcal{P}^{\alpha\beta\delta\lambda}= \frac{5  g^{\alpha\beta} g^{\delta\lambda} - g^{\alpha\lambda} g^{\beta\delta} -  g^{\alpha\delta} g^{\beta\lambda}}{72},
\end{equation*}
such that $\mathcal{P}^{\mu\nu\rho\sigma}$, $\mathcal{P}^{\mu\sigma\nu\rho}$, and $\mathcal{P}^{\mu\rho\nu\sigma}$ single out $a_1$, $a_2$, and $a_3$, respectively. The number of space--time dimensions has been set to $4$, as all the loop integrals at this order are finite.

We now come back to computing the universal coefficients. Comparing Eq.~\eqref{eq:lagr-D^4_S^2-aux} with the relevant terms from Eq.~\eqref{eq:effective-lagrangian}, we find:
\begin{align}
\label{eq:g14}
g_{14}^{ij} &= \frac{1}{2 \, n_D} \mathcal{P}^{\mu\nu\rho\sigma}  \int [d^d p] \, {\rm tr_s} \left( \slashed{\Delta}_i \slashed{\Delta}_j \gamma_\mu \slashed{\Delta}_j \gamma_\nu \slashed{\Delta}_j  \slashed{\Delta}_i \gamma_\rho \slashed{\Delta}_i \gamma_\sigma \right) \notag \\
&= -\frac{1}{6} \left( 2 m_i^2 + m_i m_j + 2 m_j^2 \right) \mathcal{I}[p^4]_{ij}^{33} + m_i^2 m_j^2 \, \mathcal{I}[p^2]_{ij}^{33} + \frac{1}{2}  m_i^3 m_j^3 \, \mathcal{I}[p^0]_{ij}^{33}, \\[1em] 
\label{eq:g15}
g_{15}^{ij} &= \frac{1}{2} g_{17}^{ij} + \frac{1}{2 \, n_D} \mathcal{P}^{\mu\sigma\nu\rho}  \int [d^d p] \, {\rm tr_s} \left( \slashed{\Delta}_i \slashed{\Delta}_j \slashed{\Delta}_i \gamma_\mu \slashed{\Delta}_i \gamma_\nu \slashed{\Delta}_i   \gamma_\rho \slashed{\Delta}_i \gamma_\sigma \right) \notag \\
&= \frac{1}{2} g_{17}^{ij} + \frac{1}{6} \mathcal{I}[p^6]_{ij}^{51} - \frac{1}{3} m_i (m_i - m_j) \mathcal{I}[p^4]_{ij}^{51} + \frac{1}{2} m_i^3 (m_i - m_j) \mathcal{I}[p^2]_{ij}^{51} + \frac{1}{2} m_i^5  m_j \mathcal{I}[p^0]_{ij}^{51}, \\[0.4em]
\label{eq:g16}
g_{16}^{ij} &= -\frac{1}{2 \, n_D} \mathcal{P}^{\mu\rho\nu\sigma}  \int [d^d p] \, {\rm tr_s} \left( \slashed{\Delta}_i \slashed{\Delta}_j \gamma_\mu \slashed{\Delta}_j \gamma_\nu \slashed{\Delta}_j  \slashed{\Delta}_i \gamma_\rho \slashed{\Delta}_i \gamma_\sigma \right) \notag \\
&= -\frac{1}{6} \left( m_i^2 - m_i m_j + m_j^2 \right) \mathcal{I}[p^4]_{ij}^{33} - \frac{1}{2} m_i m_j (m_i^2 + m_j^2) \mathcal{I}[p^2]_{ij}^{33} + \frac{1}{2}  m_i^3 m_j^3 \, \mathcal{I}[p^0]_{ij}^{33}, \\[1em]
\label{eq:g17}
g_{17}^{ij} &= -\frac{1}{2 \, n_D} \mathcal{P}^{\mu\rho\nu\sigma}  \int [d^d p] \, {\rm tr_s} \left( \slashed{\Delta}_i \slashed{\Delta}_j \gamma_\mu \slashed{\Delta}_j \slashed{\Delta}_i \gamma_\nu \slashed{\Delta}_i   \gamma_\rho \slashed{\Delta}_i \gamma_\sigma \right) \notag \\
&= -\frac{1}{4} \mathcal{I}[p^6]_{ij}^{42} + \frac{1}{3} \left( m_i^2 - m_i m_j + m_j^2 \right) \mathcal{I}[p^4]_{ij}^{42} - \frac{1}{4} m_i^2 \left( m_i^2 + 4 m_j^2 \right) \mathcal{I}[p^2]_{ij}^{42} + \frac{1}{2} m_i^4 m_j^2 \, \mathcal{I}[p^0]_{ij}^{42}.
\end{align}
Note that the $\frac{1}{2}$ factors in the expressions of $g_{14,16}^{ij}$ in Eqs.~\eqref{eq:g14} and \eqref{eq:g16} are symmetry factors multiplying the $S D^2 S D^2$ and $S D^\mu D^\nu S D_\mu D_\nu $ terms in the CDE. 

Concerning the terms involving $P$ instead of $S$, we do not write them explicitly, as they can be obtained in a straightforward manner by taking the gauge--invariant $\mathcal{O}(S^2 D^4)$ traces from Eq.~\eqref{eq:lagr-D^4_S^2-aux} (first and second lines) and substituting $S \to P$ and $g_N^{ij} \to g_N^{i(j)}$, with $N \in \{14,15,16,17\}$.

\vspace*{3mm}

\noindent\textbf{$\mathcal{O}(SP D^4)$ terms.} At dimension--6,  the $\mathcal{O}(SP D^4)$ terms are the only ones that depend  on the dual tensor $\widetilde{F}_{\mu\nu}$, therefore we treat them separately. These terms can be computed directly from the CDE in Eq.~\eqref{eq:effective-lagrangian}:
\begin{align}
\label{eq:lagr-D^4_S_P}
16 \pi^2 \mathcal{L}_{SPD^4} &= i \, c_f \left[  \int [d^d p] \, {\rm tr_s} \left( \slashed{\Delta}_i \slashed{\Delta}_j \gamma^5 \slashed{\Delta}_i \gamma_\mu \slashed{\Delta}_i \gamma_\nu \slashed{\Delta}_i   \gamma_\rho \slashed{\Delta}_i \gamma_\sigma \right) {\rm tr_g} \left( S_{ij} P_{ji} D^\mu_i D^\nu_i D^\rho_i D^\sigma_i \right) \right. \notag \\ 
&\left. + \int [d^d p] \, {\rm tr_s} \left( \slashed{\Delta}_i \gamma^5 \slashed{\Delta}_j \slashed{\Delta}_i \gamma_\mu \slashed{\Delta}_i \gamma_\nu \slashed{\Delta}_i   \gamma_\rho \slashed{\Delta}_i \gamma_\sigma \right) {\rm tr_g} \left( P_{ij} S_{ji} D^\mu_i D^\nu_i D^\rho_i D^\sigma_i \right) \right. \notag \\ 
&\left. + \int [d^d p] \, {\rm tr_s} \left( \slashed{\Delta}_i \slashed{\Delta}_j \gamma_\mu \slashed{\Delta}_j \gamma_\nu  \slashed{\Delta}_j \gamma^5  \slashed{\Delta}_i  \gamma_\rho \slashed{\Delta}_i \gamma_\sigma \right) {\rm tr_g} \left( S_{ij} D^\mu_j D^\nu_j P_{ji} D^\rho_i D^\sigma_i \right) \right] \notag \\
&= c_f \, n_D \, {\rm tr_g}  \left\{ g_{18}^{ij} \left[ \left( S_{ij} P_{ji} + P_{ij} S_{ji} \right) \widetilde{F}^{\mu\nu}_i F_{\mu\nu,i}  \right] + g_{19}^{ij} \left( S_{ij} \widetilde{F}^{\mu\nu}_j P_{ji} F_{\mu\nu,i} \right) \right\}.
\end{align}
with the universal coefficients having the rather simple expressions:
\begin{equation}
\label{eq:g18+g19}
g_{18}^{ij} = -\frac{1}{2} m_i m_j \, \mathcal{I}[p^0]_{ij}^{31}, \quad g_{19}^{ij} = -\frac{1}{2} m_i m_j \, \mathcal{I}[p^0]_{ij}^{22},
\end{equation}
and $\widetilde{F}_{\mu\nu}$ defined in Eq.~\eqref{eq:F-tilde-munu}.

The remaining two terms, $S D^\mu P D^\nu D^\rho D^\sigma$ and $P D^\mu S D^\nu D^\rho D^\sigma$, do not appear because their corresponding loop integrals vanish. This is expected from the physical point of view, as such terms could only originate from an operator such as $S_{ij} [D_\mu, S]_{ji}  [ D_\nu, \widetilde{F}^{\nu\mu} ]_i$, which vanishes due to $[ D_\nu, \widetilde{F}^{\nu\mu} ] = 0$.

\vspace*{3mm}

\noindent\textbf{$\mathcal{O}(D^6)$ terms.} The easiest way of tackling the $\mathcal{O}(D^6)$ piece is to directly compute the associated loop integral from the effective Lagrangian from Eq.~\eqref{eq:effective-lagrangian}, and then group the result in independent terms using the cyclic property of traces:
\begin{align}
\label{eq:lagr-D^6-aux}
16 \pi^2 \mathcal{L}_{D^6} &= - \frac{c_f}{6} \int [d^d p] \, {\rm tr_s} \left( \slashed{\Delta}_i \gamma_\mu \slashed{\Delta}_i \gamma_\nu \slashed{\Delta}_i   \gamma_\rho \slashed{\Delta}_i \gamma_\sigma  \slashed{\Delta}_i \gamma_\tau \slashed{\Delta}_i \gamma_\omega \right) {\rm tr_g} \left( D^\mu_i D^\nu_i D^\rho_i D^\sigma_i D^\tau_i D^\omega_i \right) \notag \\	
&= c_f  \, {\rm tr_g} \left[ - \frac{13}{90 m_i^2} \left( D^2_i D^2_i D^2_i \right) - \frac{4}{15 m_i^2} \left( D^2_i D^\mu_i  D^2_i D_{\mu,i} \right)  + \frac{17}{15 m_i^2} \left( D^2_i D^\mu_i D^\nu_i D_{\mu,i} D_{\nu,i} \right) \right. \notag \\
&\left. - \frac{3}{5 m_i^2} \left( D^\mu_i D^{\rho}_i D_{\mu,i} D^\nu_i D_{\rho,i} D_{\nu,i} \right) + \frac{1}{45 m_i^2} \left( D^\mu_i D^\nu_i D^{\rho}_i D_{\mu,i} D_{\nu,i} D_{\rho,i} \right) \right].
\end{align}
At this order, there are two possible gauge--invariant terms, which we write as:
\begin{equation}
\label{eq:lagr-D^6}
16 \pi^2 \mathcal{L}_{D^6} = c_f \, n_D \, \left[ \frac{1}{2} g_{20}^i {\rm tr_g} \left( \left[ D^\mu  , F_{\mu\nu} \right]_i \left[ D_\rho  , F^{\rho\nu} \right]_i\right) + \frac{i}{3} g_{21}^i {\rm tr_g} \left( F^{\mu}_{\;\;\: \nu} F^{\nu}_{\;\;\: \rho} F^{\rho}_{\;\;\: \mu} \right) \right].
\end{equation}
After expanding the commutators from the gauge--invariant traces and comparing Eqs.~\eqref{eq:lagr-D^6-aux} and \eqref{eq:lagr-D^6}, we are able to write down the expressions for the two universal coefficients arising at $\mathcal{O}(D^6)$:
\begin{equation}
\label{eq:g20+g21}
g_{20}^i = \frac{1}{15 m_i^2}, \quad g_{21}^i = \frac{1}{60 m_i^2}.
\end{equation}
We refrain from writing the form factors in terms of master integrals, as the expressions involved would be much lengthier. Moreover, our results for $\mathcal{O}(D^6)$ are in agreement with the ones previously obtained in Refs.~\cite{Henning:2014wua,Zhang:2016pja}.

\section{Examples} \label{sec:examples}

In this section, we apply our previously obtained results to two concrete scenarios. The first example is a toy model involving a heavy charged fermion coupling to a real pseudoscalar and a (massless) $U(1)$ gauge boson, while the second one represents a fermionic model discussed in Ref.~\cite{Angelescu:2018dkk} in the context of strongly first order electroweak phase transitions.

\subsection{A Toy Model} \label{subsec:toy-model}

The toy model that we consider contains a massless $U(1)$ gauge boson, $A_{\mu}$, and a light real pseudoscalar, $\phi$, together with a heavy charged VLF $\psi$, which is to be integrated out at one--loop. Ignoring the pieces that are irrelevant for our purposes, the Lagrangian for this toy theory reads:
\begin{equation}
\label{eq:toy-model-lagr}
\mathcal{L}_{\rm toy} \supset \overline{\psi} \left( i \gamma^\mu D_\mu - m - i  \gamma^5 \lambda \phi  \right) \psi,
\end{equation}
with $\lambda$ a real Yukawa coupling, as dictated by the hermicity of the Lagrangian. Assuming unit charge for $\psi$, the covariant derivative is given by
\begin{equation}
\label{eq:toy-model-cov-deriv}
D_\mu = \partial_\mu + i g A_\mu,
\end{equation}
where $g$ is the $U(1)$ gauge coupling. Casting the toy model Lagrangian in the form of Eq.~\eqref{eq:vlf_lagr}, we obtain the following identities:
\begin{equation}
\label{eq:toy-model-S-P}
S=0, \quad [D_\mu,S] = 0, \quad P = \lambda \phi, \quad [D_\mu,P] = \lambda \left( \partial_\mu \phi \right).
\end{equation}
Note that the covariant derivative acting on the real pseudoscalar $\phi$ reduces to the usual derivative, as expected. As there is only one heavy fermion to be integrated out, $P$ and $[D_\mu,P]$ are scalars in flavour space, which is also true for the field strength tensor, given by:
\begin{equation}
\label{eq:toy-model-Fmunu}
F_{\mu\nu} = - i \left[ D_\mu, D_\nu \right] = g \left( \partial_\mu A_\nu - \partial_\nu A_\mu  \right) \equiv g A_{\mu\nu}.
\end{equation}
Therefore, the trace in multiplet space is straightforward, and all the universal coefficients depend on only one mass scale, i.e. the heavy fermion mass $m$. The gauge trace is trivial as well, as both light fields, $\phi$ and $F_{\mu\nu}$, have no gauge quantum numbers. We now use the equal mass expressions of the universal coefficients reported in Tables~\ref{tab:univ-coeff-d-1234}--\ref{tab:univ-coeff-d-6-p2} to write down the one--loop effective Lagrangian arising from integrating out $\psi$:
\begin{align}
\label{eq:toy-model-effective-lagr}
\mathcal{L}_H^{\rm toy} &= \frac{c_f n_D}{16 \pi^2} \left[ \frac{g^2}{12} \log\frac{\mu^2}{m^2} A_{\mu\nu} A^{\mu\nu} - \frac{\lambda^2}{4} \left( 1  + \log\frac{\mu^2}{m^2} \right) \left(\partial_\mu \phi \right)^2 + \frac{\lambda^2 m^2}{4}  \left(3 + \log\frac{\mu^2}{m^2} \right) \phi^2 \right. \notag \\ 
& \left. +  \frac{\lambda^4}{4} \left( \frac{8}{3}  + \log\frac{\mu^2}{m^2} \right) \phi^4 - \frac{\lambda^6}{12 m^2} \phi^6 + \frac{5 \lambda^4}{12 m^2} \phi^2 \left(\partial_\mu \phi \right)^2 - \frac{\lambda^2}{24 m^2} \left(\partial^2 \phi \right)^2   \right. \notag \\ 
& \left. - \frac{\lambda^2 g^2}{12 m^2} \phi^2  A_{\mu\nu} A^{\mu\nu} + \frac{\lambda g^2}{4m} \phi \widetilde{A}_{\mu\nu} A^{\mu\nu} + \frac{g^3}{30 m^2} \left( \partial_\mu A^{\mu\nu} \right)^2 \right].
\end{align} 
It is worthwhile to note the absence of the $ P [D_\mu, P] [D_\nu F^{\nu\mu}]$ and $F^{\mu}_{\;\;\: \nu} F^{\nu}_{\;\;\: \rho} F^{\rho}_{\;\;\: \mu}$ terms from the results in Eq.~\eqref{eq:toy-model-effective-lagr}. The first term vanishes for real (pseudo)scalars such as $\phi$, as can be shown through integration by parts. As for the second term, one can prove that
\begin{equation}
\label{eq:furry-theorem}
{\rm tr_g} \left( F^{\mu}_{\;\;\: \nu} F^{\nu}_{\;\;\: \rho} F^{\rho}_{\;\;\: \mu} \right) = \frac{1}{2}  {\rm tr_g} \left( \left[F^{\mu}_{\;\;\: \nu}, F^{\nu}_{\;\;\: \rho} \right] F^{\rho}_{\;\;\: \mu} \right),
\end{equation}
which shows that this effective operator vanishes in the case of abelian gauge fields.

\subsection{Fermions and Cosmological Phase Transitions} \label{subsec:EWPT-fermions}

We now focus on applying the techniques from Sec.~\ref{sec:fermionic-uolea} to a more realistic model in which vector--like (VL) leptons are added to the SM particle spectrum in order to produce a strongly first order EW phase transition~\cite{Angelescu:2018dkk}. Besides the SM particle content, this model contains three VL lepton~\footnote{In the model presented in Ref.\cite{Angelescu:2018dkk}, the hypercharge of the $SU(2)_L$ doublet is $-\frac{1}{2}$, but here we keep the discussion more general and denote the doublet hypercharge as $Y$.} multiplets:
\begin{equation}
\label{eq:ewpt-vll-spectrum}
 L_{L,R} = \begin{pmatrix}
  N \\  E 
 \end{pmatrix}_{L,R}   \sim (1, 2, Y), \quad
N^\prime_{L,R} \sim  (1, 1, Y+\frac{1}{2}), \quad
 E^\prime_{L,R} \sim  (1, 1,  Y-\frac{1}{2}),
\end{equation}
where we use the notation $\left( SU(3)_C, SU(2)_L, U(1)_Y \right)$ to denote the charges of the new fermions under the SM gauge group $ SU(3)_C \times SU(2)_L \times U(1)_Y$. At the renormalizable level, the most general gauge--invariant Lagrangian involving the new fermionic fields is given by:
\begin{align}
\label{eq:ewpt-vll-lagrangian}
{\cal L}_{VLL}&= \overline{L}(  i \gamma_\mu D^\mu_L - m_L) L + \overline{E}^\prime(i \gamma_\mu D^\mu_E - m_E ) E^\prime + \overline{N}^\prime(i \gamma_\mu D^\mu_N - m_N) N^\prime  \notag \\
&-\left[ \overline{L} \, H \left( y_{E_L} \mathbb{P}_L +y_{E_R} \mathbb{P}_R \right) E^\prime + \overline{L} \, \widetilde{H} \left( y_{N_L} \mathbb{P}_L +y_{N_R} \mathbb{P}_R \right) N^\prime + \mathrm{h.c.}  \right],
\end{align}
where $\mathbb{P}_{L,R}$ are chiral projectors. The covariant derivatives acting on the fermionic fields read:
\begin{gather}
\label{eq:ewpt-vll-covar-deriv}
D_{\mu,L} = \partial_\mu + i \left( g_1 Y B_\mu + \frac{g_2}{2} W_\mu^a \sigma^a  \right), \notag \\ D_{\mu,N} = \partial_\mu + i  g_1 \left(Y + \frac{1}{2} \right) B_\mu, \quad D_{\mu,E} = \partial_\mu + i  g_1 \left(Y - \frac{1}{2} \right) B_\mu.
\end{gather}
Although not explicitly written, it is understood that the $\partial_\mu$ and $B_\mu$ pieces in $D_{\mu,L}$ are multiplied by the $2\times 2$ identity matrix in $SU(2)_L$ space. In order to match the VLL Lagrangian in Eq.~\eqref{eq:ewpt-vll-lagrangian} to the notation used in Eq.~\eqref{eq:vlf_lagr}, we define:
\begin{equation}
\label{eq:ewpt-vll-yuk-def}
y_A \equiv \frac{y_{A_L} + y_{A_R}}{2}, \quad z_A \equiv \frac{y_{A_L} - y_{A_R}}{2},
\end{equation}
with $A = E,N$. Working in the basis $\Psi = \begin{pmatrix} L & E & N \end{pmatrix}^T$, where $T$ means transposition only in flavour space, the expressions of the $S$ and $P$ matrices read:
\begin{equation}
\label{eq:ewpt-vll-S-and-P}
S = \begin{pmatrix}
0_{2 \times 2} & y_E \, H_{2 \times 1} & y_N \, \widetilde{H}_{2 \times 1} \\
y_E^* \, H^\dagger_{1 \times 2} & 0 & 0 \\
y_N^* \, \widetilde{H}^\dagger_{1 \times 2} & 0 & 0
\end{pmatrix}, \quad 
P = i \begin{pmatrix}
0_{2 \times 2} & z_E \, H_{2 \times 1} & z_N \, \widetilde{H}_{2 \times 1} \\
-z_E^* \, H^\dagger_{1 \times 2} & 0 & 0 \\
-z_N^* \, \widetilde{H}^\dagger_{1 \times 2} & 0 & 0
\end{pmatrix}.
\end{equation}
In the equation above, the subscripts denote the dimension in $SU(2)_L$ space of each element of the $S$ and $P$ matrices, while the entries that do not carry any $SU(2)_L$ indices have no subscripts. However, from now on, we stop writing the dimensions of each $SU(2)_L$ submatrix in order to simplify the notation. Concerning the mass matrix and the covariant derivative matrix, they are given by:
\begin{equation}
M = {\rm diag}( m_L , m_E, m_N), \quad D^{\mu} = {\rm diag}( D^{\mu}_L, D^{\mu}_E, D^{\mu}_N).
\end{equation}
To keep the discussion simple, we choose all VLL masses to be degenerate, $m_L = m_E = m_N \equiv m$, and work in the limit where $z_{E,N} = 0$, i.e. a vanishing $P$ matrix.~\footnote{Interestingly, achieving a strongly first--order EW phase transition in the VLL model under study favors the region of the parameter space where $z_{E,N} \simeq 0$~\cite{Angelescu:2018dkk}.} Note however that setting $P=0$ amounts to removing all CP--odd operators that might arise from integrating out the VLLs. We will return to this subject towards the end of the section.

Besides $S$, the other building blocks appearing in the computation are $\left[ D_\mu , S \right]$ and $F_{\mu\nu}$, given by:
\begin{equation}
\label{eq:ewpt-vll-Dmu-S}
\left[ D_\mu , S \right] = \begin{pmatrix}
0 & y_E \, ( D_\mu H ) & y_N \, ( D_\mu \widetilde{H} ) \\
y_E^* \, ( D_\mu H )^\dagger & 0 & 0 \\
y_N^* \, ( D_\mu \widetilde{H} )^\dagger & 0 & 0
\end{pmatrix},
\end{equation}
and
\begin{equation}
\label{eq:ewpt-vll-Fmunu}
F_{\mu\nu} = \begin{pmatrix}
g_1 Y B_{\mu\nu} + \frac{g_2}{2} W_{\mu\nu}^a \sigma^a & 0 & 0 \\ 0 & g_1 \left( Y-\frac{1}{2} \right) B_{\mu\nu} & 0 \\ 0 & 0 & g_1 \left( Y + \frac{1}{2} \right) B_{\mu\nu} \end{pmatrix}.
\end{equation}
Here, the covariant derivatives acting on the Higgs doublet are defined in Eq.~\eqref{eq:higgs-covariant-derivative} from App.~\ref{app:operator-basis}, and the $SU(2)_L$ and $U(1)_Y$ gauge field strengths follow the same notation as in Ref.~\cite{Grzadkowski:2010es}. As exemplified by Eq.~\eqref{eq:ewpt-vll-Dmu-S}, gauge invariance ensures that for any generic matrix $X(\phi)$ depending linearly on the light fields $\phi$, we have that $\left[ D_\mu , X(\phi) \right] = X(D_\mu \phi)$. 

Before listing the final result, we provide some details of our calculation. To simplify the computation, it is useful to note that $S^2$ is a diagonal matrix in flavour space:
\begin{equation}
S^2 = \begin{pmatrix}
|y_E|^2 H H^{\dagger} + |y_N|^2 \widetilde{H} \widetilde{H}^{\dagger} & 0 & 0 \\ 0 & |y_E|^2 |H|^2 & 0 \\ 0 & 0 & |y_N|^2 |H|^2 \end{pmatrix},
\end{equation}
which follows from $H^{\dagger} \widetilde{H}=0$. Similar results hold for $\left[ D_\mu , S \right]^2$ and $\left[ D_\mu, \left[ D^\mu , S \right]\right]^2$ with appropriate replacements such as $H \to (D_\mu H)$. Note however that $S \left[ D_\mu , S \right]$ is not diagonal in flavour space:
\begin{equation}
S \left[ D_\mu , S \right]= \begin{pmatrix}
|y_E|^2 H ( D_\mu H )^{\dagger} + |y_N|^2 \widetilde{H} ( D_\mu \widetilde{H} )^{\dagger} & 0 & 0 \\ 0 & |y_E|^2 H^{\dagger} ( D_\mu H ) & y_E^* y_N H^{\dagger} ( D_\mu \widetilde{H} )  \\ 0 & y_E y_N^* \widetilde{H}^{\dagger} ( D_\mu H ) & |y_N|^2 ( D_\mu H )^\dagger H \end{pmatrix}.
\end{equation}

We now write down the Lagrangian obtained after integrating out the VLLs in Eq.~\eqref{eq:ewpt-vll-spectrum} by splitting it into pieces containing $\rm dim \leq 4$ operators and another one containing $\rm dim=6$ effective operators. The first piece reads:
\begin{align}
\label{eq:eff-lagr-vll-ren}
16 \pi^2 \mathcal{L}_H^{\rm dim \leq 4} &= -\frac{1}{6} \log\frac{\mu^2}{m^2} (1 + 8 Y^2) g_1^2 B_{\mu\nu} B^{\mu\nu} -\frac{1}{6} \log\frac{\mu^2}{m^2} g_2^2 W_{\mu\nu}^a W^{\mu\nu, a} \notag \\
&+ \left(- \frac{4}{3} + 2 \log \frac{\mu^2}{m^2} \right) \left( |y_N|^2 + |y_E|^2 \right) \left| D_\mu H \right|^2 \notag \\
 &-  \left(1 + 3 \log \frac{\mu^2}{m^2} \right) \left( |y_N|^2 + |y_E|^2 \right) m^2 \left| H \right|^2 \notag \\
 &+ \left(\frac{16}{3} + 2 \log \frac{\mu^2}{m^2} \right) \left( |y_N|^4 + |y_E|^4 \right) \left| H \right|^4,
\end{align}
and renormalizes the scalar and gauge kinetic terms, plus the scalar mass and quartic terms. Note that we have used the $\overline{MS}$ scheme and dropped the divergent parts. Using the SMEFT operator basis defined in App.~\ref{app:operator-basis}, the (CP--even) dimension--6 effective Lagrangian generated by integrating out  at one--loop the VLLs in our model is given by:
\begin{align}
\label{eq:eff-lagr-vll-cp-even}
16 \pi^2 \mathcal{L}_H^{\rm CP} &= -\frac{2 \, (|y_N|^6 + |y_E|^6)}{15 m^2} \mathcal{O}_6 + \frac{4(|y_N|^2 + |y_E|^2)^2}{5 m^2} \mathcal{O}_H - \frac{4 (|y_N|^2 - |y_E|^2)^2}{5m^2} \mathcal{O}_T \notag  \\
&-  \frac{2 \left( |y_N|^4 - 4 |y_N|^2 |y_E|^2 + |y_E|^4\right)}{5 m^2} \mathcal{O}_f   -\frac{7 \left(|y_N|^2+|y_E|^2\right)}{120 m^2} \mathcal{O}_{WW} \notag \\ 
&- \frac{ \left(7  + 40 Y + 80 Y^2 \right) |y_N|^2 + \left(7  - 40 Y + 80 Y^2 \right) |y_E|^2}{120 m^2}\mathcal{O}_{BB} \notag \\ 
&+ \frac{(3+20 Y) |y_N|^2+(3 - 20 Y) |y_E|^2}{60 m^2} \mathcal{O}_{WB}  + \frac{|y_N|^2+|y_E|^2}{5 m^2} \mathcal{O}_{K4} \notag \\ 
&- \frac{4 \left(|y_N|^2 + |y_E|^2\right)}{15 m^2} \mathcal{O}_B - \frac{4 \left(|y_N|^2 + |y_E|^2\right)}{15 m^2} \mathcal{O}_W + \frac{ 2 + 16 Y^2 }{15 m^2} \mathcal{O}_{2B}  \notag \\ 
&+\frac{2}{15 m^2} \mathcal{O}_{2W}  + \frac{1}{30 m^2} \mathcal{O}_{3W}.
\end{align}
There are several simple consistency checks that one can perform to assess the validity of the results presented in Eq.~\eqref{eq:eff-lagr-vll-cp-even}. For example, the leading contribution to the $T$ parameter~\cite{Peskin:1991sw,Altarelli:1990zd} is proportional to the Wilson coefficient of $\mathcal{O}_T$:
\begin{equation}
\label{eq:T-parameter}
4 \pi e^2 T \simeq \frac{2 (|y_N|^2 - |y_E|^2)^2 v^2 }{5m^2},
\end{equation}
and vanishes in the custodial limit $y_E = y_N$, as expected. In the above equation, $e$ is the electromagnetic coupling constant and $v$ the Higgs VEV. We have explicitly checked that our expression for the $T$ parameter from Eq.~\eqref{eq:T-parameter} matches the one obtained in Ref.~\cite{Ellis:2014dza}. Furthermore, the Wilson coefficient of $\mathcal{O}_6$ can alternatively be computed from the Coleman--Weinberg potential~\cite{Coleman:1973jx} corresponding to the fermionic Lagrangian in Eq.~\eqref{eq:ewpt-vll-lagrangian}, and we have explicitly checked that the two methods give the same result. 

Yet another consistency check can be done by inspecting the physical Higgs boson's loop--induced coupling to photons. The leading $\mathcal{O}(m^{-2})$ VLL contribution to the $h\gamma\gamma$ coupling can be derived from Eq.~\eqref{eq:eff-lagr-vll-cp-even} as well as through low--energy Higgs theorems~\cite{Ellis:1975ap,Shifman:1979eb} (see also Refs.~\cite{Carena:2012xa,Joglekar:2012vc,Voloshin:2012tv,Bizot:2015zaa} for VL fermion applications of low--energy Higgs theorems). Denoting the Wilson coefficient of a given operator $\mathcal{O}_X$ as $C_X$, the VLL contribution to the $h\gamma\gamma$ coupling can be read from:
\begin{equation}
\label{eq:ewpt-vll-h-gamma-gamma}
\mathcal{L}_H^{\rm CP} \supset (C_{BB} + C_{WW} - C_{WB}) \, e^2 |H|^2 A_{\mu\nu}A^{\mu\nu} \supset - \frac{e^2 Q_E^2}{24 \pi^2} \frac{v   |y_E|^2 }{m^2}  h A_{\mu\nu}A^{\mu\nu},
\end{equation}
where $e$ is the electromagnetic coupling constant, $A_{\mu\nu}$ the electromagnetic field strength, and $Q_E = Y - \frac{1}{2}$ is the electrical charge of $E^{(\prime)}$. For simplicity, we have set $y_N=0$ in Eq.~\eqref{eq:ewpt-vll-h-gamma-gamma}, as the matching for  $y_N \neq 0$ proceeds in an analogous way. Working in the unitary gauge, the Higgs doublet becomes 
\begin{equation*}
H \to \frac{v+h}{\sqrt{2}} \begin{pmatrix} 0 \\ 1 \end{pmatrix}
\end{equation*}  
upon electroweak symmetry breaking, with $v$ the vacuum expectation value and $h$ the physical Higgs scalar. 

Using the low--energy theorems (LET), the (CP--even) effective $h\gamma\gamma$ coupling arising from integrating out the heavy VL leptons from Eq.~\eqref{eq:ewpt-vll-spectrum} is described by~\cite{Bizot:2015zaa}
\begin{equation}
\label{eq:low-energy-th-CP}
\mathcal{L}_{h\gamma\gamma}^{\rm LET, \, CP} = \frac{e^2}{48 \pi^2} Q_E^2 \frac{\partial}{\partial v} \log \left[ \det \left( \mathcal{M}_E^{\dagger} \mathcal{M}_E \right) \right] h A_{\mu\nu}A^{\mu\nu},
\end{equation}
where $\mathcal{M}_E$ is the $ E - E^\prime $ mass matrix and, again, we have set $y_N = 0$ for simplicity. The $E$--sector mass matrix is extracted from Eq.~\eqref{eq:ewpt-vll-lagrangian} and reads:
\begin{equation}
\label{eq:mass-matrix}
\mathcal{M}_E = \begin{pmatrix} m & \frac{v}{\sqrt{2}} y_{E_R} \\ \frac{v}{\sqrt{2}} y_{E_L}^* & m  \end{pmatrix} \xrightarrow{y_{E_L} = y_{E_R} \equiv y_E} \begin{pmatrix} m & \frac{v}{\sqrt{2}} y_E \\ \frac{v}{\sqrt{2}} y_E^* & m  \end{pmatrix}.
\end{equation}
Plugging $\mathcal{M}_E$ in Eq.~\eqref{eq:low-energy-th-CP} and expanding up to $\mathcal{O}(m^{-2})$, we find the same result as in Eq.~\eqref{eq:ewpt-vll-h-gamma-gamma}, which constitutes a useful check for our results in Eq.~\eqref{eq:eff-lagr-vll-cp-even}.

Using methods similar to Ref.~\cite{Henning:2014wua}, Ref.~\cite{Huo:2015exa} performed the same computation as in our Eq.~\eqref{eq:eff-lagr-vll-cp-even}. However, our results differ from theirs even after taking into account the redundancies in the operator basis used in Ref.~\cite{Huo:2015exa}.~\footnote{After translating the results of Ref.~\cite{Huo:2015exa} into the operator basis used in Eq.~\eqref{eq:eff-lagr-vll-cp-even}, we find agreement only for the Wilson coefficients of $\mathcal{O}_6$, $\mathcal{O}_T$, and $\mathcal{O}_{GG}$.}
	
As advertised earlier, we now turn our attention towards CP--odd dimension--6 operators, which can be generated only if $P \neq 0$. Therefore, in the following, we consider $z_{E,N} \neq 0$, and focus only on the CP--odd effective Lagrangian. In the VLL model at hand, the CP--odd operators arise from the $g_{13}^{ij(k)(l)}, \, g_{13}^{i(j)(k)l}$ terms in Eq.~\eqref{eq:lagr-X^4_D^2} and from the $\mathcal{O}(SPD^4)$ Lagrangian in Eq.~\eqref{eq:lagr-D^4_S_P}. Putting together all these contributions, we find:
\begin{align}
\label{eq:eff-lagr-vll-cp-odd}
16 \pi^2 \mathcal{L}_H^{\cancel{\rm CP}} &= - \frac{ ( |y_{N_L}|^2 + |y_{N_R}|^2 ) {\rm Im}(y_{N_L} y_{N_R}^*) - ( |y_{E_L}|^2 + |y_{E_R}|^2 ) {\rm Im}(y_{E_L} y_{E_R}^*)}{3 m^2} \widetilde{\mathcal{O}}_f \notag \\
& - \frac{ \left( 1 + 6Y + 12Y^2 \right){\rm Im}(y_{N_L} y_{N_R}^*) + \left( 1 - 6Y + 12Y^2 \right){\rm Im}(y_{E_L} y_{E_R}^*)}{12 m^2} \widetilde{\mathcal{O}}_{BB} \notag \\
& + \frac{ \left( 1 + 6Y \right){\rm Im}(y_{N_L} y_{N_R}^*) + \left( 1 - 6Y \right){\rm Im}(y_{E_L} y_{E_R}^*)}{12 m^2} \widetilde{\mathcal{O}}_{WB} \notag \\
& - \frac{{\rm Im}(y_{N_L} y_{N_R}^* + y_{E_L} y_{E_R}^*)}{12 m^2} \widetilde{\mathcal{O}}_{WW},
\end{align}
where we have used $y_{A_{L,R}}$ instead of $y_A$ and $z_A$, cf. Eq.~\eqref{eq:ewpt-vll-yuk-def}. Since $\widetilde{\mathcal{O}}_f$ breaks custodial symmetry, its Wilson coefficient vanishes when the VLL sector respects custodial symmetry, which is a helpful check of the results in Eq.~\eqref{eq:eff-lagr-vll-cp-odd}. Another validity check is to compute the CP--odd $h\gamma\gamma$ effective coupling from Eq.~\eqref{eq:eff-lagr-vll-cp-odd} and compare it with the result obtained from low--energy theorems. The comparison carries on similarly to the CP--even case discussed earlier. Starting from Eq.~\eqref{eq:eff-lagr-vll-cp-odd}, we isolate the relevant part of the CP--odd Lagrangian:
\begin{equation}
\label{eq:ewpt-vll-h-gamma-gamma-cp-odd}
\mathcal{L}_H^{\cancel{\rm CP}} \supset - \frac{e^2 Q_E^2}{16 \pi^2} \frac{v \, {\rm Im}(y_{E_L} y_{E_R}^*)}{m^2}  h \widetilde{A}_{\mu\nu}A^{\mu\nu},
\end{equation}
where again we have set $y_{N_{L,R}} = 0$. Using the low--energy theorem instead, the VLL contribution to the CP--odd effective $h\gamma\gamma$ coupling~\footnote{Note that the minus sign in Eq.~\eqref{eq:low-energy-th-CP-odd} is usually omitted in the literature, as it has no physical impact.} reads:
\begin{equation}
\label{eq:low-energy-th-CP-odd}
\mathcal{L}_{h\gamma\gamma}^{\rm LET, \,{\cancel{\rm CP}}} = -\frac{e^2}{16 \pi^2} Q_E^2 \frac{\partial}{\partial v} {\rm arg} \left[ \det \left(\mathcal{M}_E \right) \right] h \widetilde{A}_{\mu\nu}A^{\mu\nu} \simeq -\frac{e^2 Q_E^2}{16 \pi^2} \frac{v \, {\rm Im}(y_{E_L} y_{E_R}^*)}{m^2}  h \widetilde{A}_{\mu\nu}A^{\mu\nu},
\end{equation}
where we have used the general expression for $\mathcal{M}_E$, given in Eq.~\eqref{eq:mass-matrix}, and kept only the leading $\mathcal{O}(m^{-2})$ contribution. Again, we find agreement between our result and the LET result.

Finally, let us comment on how our results change if the fermions from Eq.~\eqref{eq:ewpt-vll-spectrum} would carry $SU(3)_C$ color charge. We assume that all three fermionic multiplets belong to the same $SU(3)_C$ representation $R$, since otherwise gauge invariance would prevent them from coupling to the SM Higgs doublet. In this case, all the Wilson coefficients of the operators listed in Eqs.~(\ref{eq:eff-lagr-vll-ren}, \ref{eq:eff-lagr-vll-cp-even}, \ref{eq:eff-lagr-vll-cp-odd}) should be multiplied by the dimension of the respective representation, $d(R)$ (e.g. 3 for the fundamental representation of $SU(3)_C$ and 8 for the adjoint). Operators involving gluons would also be generated, the CP--even ones being given by:
\begin{align}
\label{eq:ewpt-vll-gluonic-cp-even}
16 \pi^2 \mathcal{L}_H^{\rm gluon, \, CP} &= -\frac{4}{3} \log \frac{\mu^2}{m^2} \, T(R) g_3^2 G_{\mu\nu}^A G^{\mu\nu,A} - \frac{2 ( |y_N|^2 + |y_E|^2 )}{3m^2} T(R) \mathcal{O}_{GG} \notag \\
& + \frac{16}{15 m^2} T(R) \mathcal{O}_{2G} + \frac{4}{15 m^2} T(R) \mathcal{O}_{3G},
\end{align}
where $g_3$ is the $SU(3)_C$ coupling constant and $T(R)$ is the Dynkin index of the $SU(3)_C$ representation $R$ to which the heavy fermions belong. The gluonic operators $\mathcal{O}_{GG}$, $\mathcal{O}_{2G}$, and $\mathcal{O}_{3G}$ are defined in App.~\ref{app:operator-basis}. The Dynkin index is defined by ${\rm tr_g} \left( t^A_R t^B_R \right) = T(R) \delta^{AB}$, with $t^A_R$ the generators in the $R$ representation of $SU(3)_C$. For instance, $T(\mathbf{3})= \frac{1}{2}$ and $T(\mathbf{8})= 3$ for the fundamental and adjoint representations of $SU(3)_C$, respectively. There is also one CP--odd gluonic operator generated:
\begin{equation}
\label{eq:ewpt-vll-gluonic-cp-odd}
16 \pi^2 \mathcal{L}_H^{\rm gluon, \, {\cancel{\rm CP}}} = - \frac{{\rm Im}(y_{N_L} y_{N_R}^* + y_{E_L} y_{E_R}^*)}{m^2} T(R) \widetilde{\mathcal{O}}_{GG},
\end{equation}
with $\widetilde{\mathcal{O}}_{GG}$ defined in App.~\ref{app:operator-basis}.

\section{Summary and Conclusions}
\label{sec:conclusions}

Initially proposed more than 30 years ago~\cite{Gaillard:1985uh,Cheyette:1987qz}, functional methods for integrating out heavy degrees of freedom have been recently revived in Ref.~\cite{Henning:2014wua} and are currently an ongoing scientific effort~\cite{Drozd:2015rsp,delAguila:2016zcb,Boggia:2016asg,Henning:2016lyp,Ellis:2016enq,Fuentes-Martin:2016uol,Zhang:2016pja,Ellis:2017jns}. Although interesting from the theoretical point of view, the true power of functional methods lies in their practical applications: once a UV sector is specified, matching at one--loop to a low--energy EFT can be achieved by using a few universal master formulas and evaluating matrix traces in internal space (e.g. spin, gauge, flavour).

In this paper, we have extended the universal one--loop formulas presented in Refs.~\cite{Henning:2014wua,Drozd:2015rsp,Henning:2016lyp,Ellis:2016enq,Fuentes-Martin:2016uol,Zhang:2016pja,Ellis:2017jns} to include the case of heavy vector--like fermions whose left and right chiralities have different Yukawa interactions with light scalars (e.g. the Higgs boson), as encoded by the $S$ and $P$ matrices in Eq.~\eqref{eq:vlf_lagr}. We have considered the limit where the new fermions do not mix with the SM fermions, the resulting universal coefficients being referred to in the literature as ``heavy--only'' coefficients. The computations have been performed in the unbroken phase, such that there are no vector and axial current interaction terms for the new fermions (besides the ones encoded in the covariant derivative matrix). 

Interestingly, as exemplified throughout Sec.~\ref{sec:fermionic-uolea}, the coefficients of operators containing an even number of $P$ insertions can be easily computed from the corresponding $S$--only coefficients by flipping the signs of one or more masses and appropriately adjusting the symmetry factors. This led to a drastically simpler computation: out of the 44 universal coefficients arising at dimension--5 and dimension--6, we needed to calculate only 15. The most striking examples are the operators arising at $\mathcal{O}(X^6)$ and $\mathcal{O}(X^4 D^2)$. For the former, we had to compute just one coefficient instead of eight, whereas for the latter we had to compute just two coefficients instead of fourteen. As an exception to this rule, in the case of dimension--4 or less coefficients we also had to add a finite correction, which stems from the BMHV treatment of $\gamma^5$ matrices in $d$ dimensions.

All in all, we find that the heavy-only fermionic UOLEA computed in the unbroken phase is described by 56 independent operators and their corresponding coefficients, which we summarize in Tables~\ref{tab:univ-coeff-d-1234}--\ref{tab:univ-coeff-d-6-p2} in App.~\ref{app:equal-masses}. As a cross--check of our computation, we have computed the $S$--only coefficients using the universal master formula from Refs.~\cite{Drozd:2015rsp,Zhang:2016pja} and found agreement between the two methods (a ``dictionary'' between the fermionic universal coefficients and the bosonic UOLEA coefficients is provided in App.~\ref{app:uolea-coeffs}). Finally, we have applied our results in Sec.~\ref{sec:examples} and integrated out heavy fermions (i) in a toy model with a pseudoscalar Yukawa interaction and (ii) in a more realistic fermionic model which can accommodate a strongly first--order electroweak phase transition.

Concerning future directions, it would certainly be useful to take advantage of the universality of functional methods and derive more general one--loop master formulas that would cover cases involving e.g. open covariant derivatives or mixed statistics (i.e. heavy bosons and fermions integrated out simultaneously). Such developments would considerably simplify phenomenological studies of a broad class of New Physics scenarios, as performing the one--loop matching to e.g. the SMEFT would translate to using a few universal formula(s) and calculating algebraic traces, which also opens up the potential for automation.

\section*{Acknowledgments}
\label{sec:acknowledgments}
While this study was ongoing, we became aware of Ref.~\cite{fermion-uolea}, which discusses similar ideas. We would like to thank the authors of Ref.~\cite{fermion-uolea} for exchanges on their related work. This work is supported by University of Nebraska-Lincoln, National Science Foundation under grant number PHY-1820891, and the NSF Nebraska EPSCoR under grant number OIA-1557417.

\appendix

\renewcommand{\theequation}{\thesection.\arabic{equation}}

\section{Master Integrals and Treatment of $\gamma^5$} \label{app:master-integrals}
\setcounter{equation}{0}
In this appendix, we provide a general definition for the master integrals used to calculate the fermionic one--loop coefficients, and then discuss our treatment of $\gamma^5$ in $d$--dimensional loop integrals. We define the master integrals as:
\begin{equation}
\label{eq:master-integrals}
\int [d^d p] \,  p^{2 n_p} \left(\Delta_i^{n_i} \Delta_j^{n_j} \cdots  \right)   \equiv \mathcal{I}[p^{2 n_p}]^{n_i n_j \cdots}_{ij \cdots}, \quad \Delta_i \equiv \frac{1}{p^2 - m_i^2},
\end{equation}
from which one can derive the following relation, which connects our definition to the one in Ref.~\cite{Zhang:2016pja}:
\begin{equation}
\label{eq:master-integrals-2}
\int [d^d p] \left( p_{\mu_1} \cdots p_{\mu_{2 n_p}} \right) \left(\Delta_i^{n_i} \Delta_j^{n_j} \cdots \right)   = \frac{\Gamma(\frac{d}{2})}{2^{n_p} \Gamma(\frac{d}{2} + n_p)} g_{\mu_1 \cdots \mu_{2 n_p}} \mathcal{I}[p^{2 n_p}]^{n_i n_j \cdots}_{ij \cdots},
\end{equation}
with $g_{\mu_1 \cdots \mu_{2 n_p}}$ the completely symmetric tensor, i.e. $g_{\mu\nu\rho\sigma} = g_{\mu\nu} g_{\rho\sigma} + g_{\mu\rho} g_{\nu\sigma} + g_{\mu\sigma} g_{\nu\rho}$ for $n_p =2$, and $d = 4 - 2 \epsilon$ the number of space--time dimensions. Throughout the paper, we use Package--X~\cite{Patel:2015tea,Patel:2016fam} to compute the fermion traces and the resulting master integrals. Analytical expressions of the master integrals can be found in Ref.~\cite{Zhang:2016pja}.

We now focus on the problems that arise when dealing with the $\gamma^5$ matrix in $d=4-2\epsilon$ dimensions. Alongside the Levi--Civita tensor, $\varepsilon_{\mu\nu\rho\sigma}$, $\gamma^5$  is an intrinsically 4--dimensional object and therefore nontrivial to define in $d$ dimensions. In this work, we address this issue by adopting the ``Breitenlohner--Maison--’t Hooft--Veltman" (BMHV) scheme~\cite{tHooft:1972tcz,Breitenlohner:1977hr}, in which the $d$--dimensional space is formally separated into a direct sum between a 4--dimensional and a $(-2\epsilon)$--dimensional subspace. As a result, each Lorentz vector/tensor now possesses  4--dimensional and $(-2\epsilon)$--dimensional components. Following Ref.~\cite{Belusca-Maito:2020ala}, we denote the former components by a bar, and the latter by a hat. For example, the $d$--dimensional metric $g_{\mu\nu}$ is written as: 
\begin{equation}
\label{eq:app-bmhv-1}
g_{\mu\nu} = \bar{g}_{\mu\nu} + \hat{g}_{\mu\nu},
\end{equation}
where $\bar{g}_{\mu\nu}$ and $\hat{g}_{\mu\nu}$ reside in 4--dimensional and $(-2\epsilon)$--dimensional spaces, respectively. They act as projectors onto these spaces, i.e.
\begin{equation}
\label{eq:app-bmhv-2}
\bar{a}_{\mu} = \bar{g}_{\mu\nu} a^\nu, \quad \hat{a}_{\mu} = \hat{g}_{\mu\nu} a^\nu, \quad \bar{a}_{\mu\nu} = \bar{g}_{\mu\rho} a^\rho_{\;\;\nu}, \quad \hat{a}_{\mu\nu} = \hat{g}_{\mu\rho}  a^\rho_{\;\;\nu},
\end{equation}
for a generic Lorentz vector/tensor $a$, and obey the following properties:
\begin{equation}
\label{eq:app-bmhv-3}
g_{\mu\nu} g^{\mu\nu} = d, \quad \bar{g}_{\mu\nu} \bar{g}^{\mu\nu} =4, \quad \hat{g}_{\mu\nu} \hat{g}^{\mu\nu} = d-4, \quad \bar{g}_{\mu\nu} \hat{g}^{\mu\nu} = 0.
\end{equation}
For a full list of properties, we refer the reader to Ref.~\cite{Belusca-Maito:2020ala}. In the following, we just list a few selected relations that are useful for our purposes. We note that relation~\eqref{eq:app-bmhv-2} applies to Dirac matrices too, from which we can deduce some of the following properties:
\begin{equation}
\label{eq:app-bmhv-4}
\{ \gamma^\mu, \gamma^\nu \} = 2 g^{\mu\nu} \mathbb{1}, \quad \{ \hat \gamma^\mu, \hat \gamma^\nu \} = 2 \hat g^{\mu\nu} \mathbb{1}, \quad \gamma_\mu \gamma^\mu = d \mathbb{1}, \quad \hat \gamma_\mu \hat \gamma^\mu = (d-4) \mathbb{1},
\end{equation}
where $\mathbb{1}$ is the identity matrix in spinor space. Finally, $\gamma^5$ no longer anticommutes with $\gamma_\mu$, but retains	some of its usual 4D properties, such as squaring to identity:
\begin{equation}
\label{eq:app-bmhv-5}
\left\{ \gamma^5, \gamma^\mu \right\} = \left\{ \gamma^5, \hat \gamma^\mu \right\} = 2 \gamma^5 \hat \gamma^\mu, \quad \left[\gamma^5, \hat \gamma^\mu \right] = 0, \quad \left( \gamma^5 \right)^2 = \mathbb{1}.
\end{equation}

As an example, let us calculate the finite correction $\delta g_2^{ij}$ appearing in Sec.~\ref{subsec:dim2}. We start by noting that Eq.~\eqref{eq:app-bmhv-5} implies
\begin{equation}
i \gamma^5 (\slashed{p} + m) i \gamma^5 = (\slashed{p} - m) - 2 \hat p_\mu \gamma^\mu,
\end{equation}
from which it follows that
\begin{align}
{\rm tr_s} \left[ \slashed{\Delta}_i \left( i \gamma^5 \right) \slashed{\Delta}_j \left(i \gamma^5 \right) \right] &= \Delta_i \Delta_j  \left\lbrace {\rm tr_s} \left[ (\slashed{p} + m_i) (\slashed{p} - m_j) \right] - 2 \hat p_\mu {\rm tr_s} \left[ (\slashed{p} + m_i) \gamma^\mu  \right] \right\rbrace \notag \\[0.4em]
& \supset - 2 n_D \Delta_i \Delta_j \hat{g}_{\mu\nu} p^\mu p^\nu. 
\end{align}
where $\Delta_{i,j}$ is defined in Eq.~\eqref{eq:master-integrals}, and we have omitted the contribution that does not vanish in 4--dimensional space, as it is not relevant for the computation of $\delta g_2^{ij}$. Using the equation above, we calculate $\delta g_2^{ij}$ as:
\begin{equation}
\delta g_2^{ij} = - 2 \hat{g}_{\mu\nu} \int [d^d p] \, p^\mu p^\nu \Delta_i \Delta_j = - \frac{2(d-4)}{d} \mathcal{I} [p^2]_{ij}^{11} = m_i^2 + m_j^2,
\end{equation}
where we have replaced $p^\mu p^\nu \to \frac{1}{d} g^{\mu\nu} p^2$ under the integral, used $\hat{g}_{\mu\nu} g^{\mu\nu} = \hat{g}_{\mu\nu} \hat{g}^{\mu\nu} = d-4$, and then took the limit $\epsilon \to 0$ in the last step.

\section{Operator Basis and $SU(2)$ Traces} \label{app:operator-basis}
\setcounter{equation}{0}

In this appendix, we define the SMEFT dimension--6 operator basis used in Sec.~\ref{subsec:EWPT-fermions}, and provide the expressions for the $SU(2)$ gauge traces encountered in Sec.~\ref{subsec:EWPT-fermions}. We start by listing the CP--even operators appearing in our computation:
\begin{align}
\label{eq:CP-even-oper}
\mathcal{O}_6 &= |H|^6, \quad \mathcal{O}_H = \frac{1}{2} \left( \partial_\mu |H|^2 \right)^2, \quad \mathcal{O}_T = | H^{\dagger} D_\mu H|^2, \quad \mathcal{O}_f = \frac{1}{2} |H|^2  \left( H^{\dagger} D^2 H + {\rm h.c.} \right), \notag \\
\mathcal{O}_{BB} &= g_1^2 |H|^2 B_{\mu\nu} B^{\mu\nu}, \quad \mathcal{O}_{WB} = g_1 g_2 (H^\dagger \sigma^a H) W_{\mu\nu}^a B^{\mu\nu}, \quad \mathcal{O}_{WW} = g_2^2 |H|^2 W_{\mu\nu}^a W^{\mu\nu,a}, \notag \\
\mathcal{O}_B &= \frac{i g_1}{2} \left(H^{\dagger} \overset{\leftrightarrow}{D_\mu} H \right) \partial_\nu B^{\mu\nu}, \quad \mathcal{O}_W = \frac{i g_2}{2} \Big(H^{\dagger} \overset{\leftrightarrow}{D^a_\mu} H \Big) \left( D_\nu W^{\mu\nu} \right)^a, \quad \mathcal{O}_{K4} = \left| D^2 H \right|^2, \notag \\
\mathcal{O}_{2B} &= -\frac{g_1^2}{2} \left( \partial^\mu B_{\mu\nu} \right)^2, \quad \mathcal{O}_{2W} = -\frac{g_2^2}{2} \left[ \left( D^\mu W_{\mu\nu} \right)^a \right]^2, \quad \mathcal{O}_{3W} = \frac{g_2^3}{6} \epsilon^{abc} W_\mu^{a\nu} W_\nu^{b\rho} W_\rho^{c\mu}, \notag \\
\mathcal{O}_{GG} &= g_3^2 |H|^2 G_{\mu\nu}^A G^{\mu\nu,A}, \quad \mathcal{O}_{2G} = -\frac{g_3^2}{2} \left[ \left( D^\mu G_{\mu\nu} \right)^A \right]^2, \quad \mathcal{O}_{3G} = \frac{g_3^3}{6} f^{ABC} G_\mu^{A\nu} G_\nu^{B\rho} G_\rho^{C\mu},
\end{align}
where  $g_{1,2.3}$ are the $U(1)_Y$, $SU(2)_L$, and $SU(3)_C$ gauge couplings, respectively, and $\sigma^{a=1,2,3}$ are the regular Pauli matrices. The structure constants of $SU(2)_L$ and $SU(3)_C$ are represented as $\epsilon^{abc}$ and $f^{ABC}$, respectively. The following notation was used:
\begin{equation}
\label{eq:left-right-derivative}
H^{\dagger} \overset{\leftrightarrow}{D_\mu} H \equiv H^{\dagger}\left( D_\mu - \overset{\leftarrow}{D_\mu} \right) H, \quad H^{\dagger} \overset{\leftrightarrow}{D^a_\mu} H \equiv H^{\dagger} \left( \sigma^a D_\mu - \overset{\leftarrow}{D_\mu} \sigma^a \right)  H,
\end{equation}
while the definitions of the field strength tensors	and their covariant derivatives are the same as in Ref.~\cite{Grzadkowski:2010es}. In addition to the 16 CP--even operators, we also encounter 5 CP--odd operators, which we define as follows:
\begin{gather}
\label{eq:CP-odd-oper}
\widetilde{\mathcal{O}}_f = \frac{i}{2} |H|^2  \left[ (D^2 H)^{\dagger} H - H^{\dagger} D^2 H  \right], \quad \widetilde{\mathcal{O}}_{WB} = g_1 g_2 (H^\dagger \sigma^a H) \widetilde{W}_{\mu\nu}^a B^{\mu\nu},  \notag \\
\widetilde{\mathcal{O}}_{BB} = g_1^2 |H|^2 \widetilde{B}_{\mu\nu} B^{\mu\nu}, \quad \widetilde{\mathcal{O}}_{WW} = g_2^2 |H|^2 \widetilde{W}_{\mu\nu}^a W^{\mu\nu,a}, \quad \widetilde{\mathcal{O}}_{GG} = g_3^2 |H|^2 \widetilde{G}_{\mu\nu}^A G^{\mu\nu,A}
\end{gather}
with $\widetilde{B}_{\mu\nu}$, $\widetilde{W}_{\mu\nu}^a$, and $\widetilde{G}_{\mu\nu}^A$ defined as in Eq.~\eqref{eq:F-tilde-munu}.

We now discuss the $SU(2)$ gauge traces appearing in our computation. The building blocks of these traces are the Higgs doublet $H$, its conjugate $ \widetilde{H} = i \sigma^2 H^{*}$, and their covariant derivatives, which read: 
\begin{equation}
\label{eq:higgs-covariant-derivative}
D_\mu H = \left(\partial_\mu + i \frac{g_1}{2}  B_\mu + i \frac{g_2}{2}  W_\mu^a \sigma^a  \right) H, \quad D_\mu \widetilde{H} = \left(\partial_\mu - i \frac{g_1}{2}  B_\mu + i \frac{g_2}{2}  W_\mu^a \sigma^a  \right) \widetilde{H},
\end{equation}
where the hypercharge of the Higgs doublet (conjugate Higgs doublet) is $\frac{1}{2}$ ($-\frac{1}{2}$). Gauge invariance ensures that traces with an odd number of $H$'s or $\widetilde{H}$'s (with or without covariant derivatives) vanish. Moreover, further simplifications are possible with the help of the following identities:
\begin{gather}
\label{eq:h-doublet-identities}
H^\dagger \widetilde{H} = (D_\mu H)^\dagger (D^\mu \widetilde{H}) = 0, \quad \widetilde{H}^\dagger \widetilde{H} = H^\dagger H \equiv |H|^2, \quad \widetilde{H}^{\dagger} (D_\mu \widetilde{H}) = (D_\mu H)^{\dagger} H \notag \\ 
(D_\mu \widetilde{H})^\dagger (D^\mu \widetilde{H}) = (D_\mu H)^\dagger (D^\mu H) \equiv |D_\mu H|^2,
\end{gather}
which follow from the definition of $\widetilde{H}$ and from  $ (D_\mu \widetilde{H}) = i \sigma^2 (D_\mu H)^*$. With all this in mind, we list the expressions of several gauge traces appearing throughout our computation:
\begin{align}
{\rm tr_g} \left[ H (D_\mu H)^{\dagger} H (D^\mu H)^{\dagger}   \right] &= {\rm tr_g} \left[ \widetilde{H}^{\dagger} (D_\mu \widetilde{H}) \widetilde{H}^{\dagger} (D_\mu \widetilde{H})\right]  = - \mathcal{O}_T + \mathcal{O}_H + i \widetilde{\mathcal{O}}_f,  \\
{\rm tr_g} \left[ H^{\dagger} (D_\mu H) H^{\dagger} (D_\mu H)\right]  &= {\rm tr_g} \left[ \widetilde{H} (D_\mu \widetilde{H})^{\dagger} \widetilde{H} (D^\mu \widetilde{H})^{\dagger}   \right] =  - \mathcal{O}_T + \mathcal{O}_H - i \widetilde{\mathcal{O}}_f, \\
{\rm tr_g} \left[ H (D_\mu H)^{\dagger} \widetilde{H} (D^\mu \widetilde{H})^{\dagger}   \right] &= {\rm tr_g} \left[ H^{\dagger} (D_\mu \widetilde{H}) \widetilde{H}^{\dagger} (D_\mu H)\right]  =  \mathcal{O}_T + \mathcal{O}_H + \mathcal{O}_f, \\
{\rm tr_g} \left[ H H^{\dagger} (D_\mu H) (D^\mu H)^{\dagger} \right] &= {\rm tr_g}  \left[ \widetilde{H} \widetilde{H}^{\dagger} (D_\mu \widetilde{H}) (D^\mu \widetilde{H})^{\dagger} \right] = \mathcal{O}_T, \\
{\rm tr_g} \left[ H H^{\dagger} (D_\mu \widetilde{H}) (D^\mu \widetilde{H})^{\dagger} \right] &=  {\rm tr_g} \left[ \widetilde{H} \widetilde{H}^{\dagger} (D_\mu H) (D^\mu H)^{\dagger} \right] = -\mathcal{O}_T -\mathcal{O}_H - \mathcal{O}_f, \\
{\rm tr_g} \left[ H^{\dagger} H (D^\mu H)^{\dagger} (D_\mu H) \right] &= - \mathcal{O}_H - \mathcal{O}_f.
\end{align}
Other traces that are not listed above follow from the cyclic property, by applying the identities in Eq.~\eqref{eq:h-doublet-identities}, or are trivial to compute, such as the traces containing only $H$ and $\widetilde{H}$ and no covariant derivatives. For completeness, we also list some (well--known) identities involving the Pauli matrices, which are useful in computing operators involving $SU(2)_L$ field strength tensors:
\begin{gather}
\widetilde{H}^{\dagger} \sigma^a \widetilde{H} = - H^{\dagger} \sigma^a H, \quad \widetilde{H}^{\dagger} \sigma^a (D_\mu\widetilde{H}) = - (D_\mu H)^{\dagger} \sigma^a H, \notag \\
 \frac{1}{2} \left[ \sigma^a, \sigma^b \right] = i \epsilon^{abc} \sigma^c, \quad \frac{1}{2} \left\{ \sigma^a, \sigma^b \right\} = \delta^{ab}.
\end{gather}

\section{Relation to the UOLEA Coefficients} \label{app:uolea-coeffs}
\setcounter{equation}{0}

We now provide the relations between our coefficients, $g_N$, to the (symmetrized) bosonic UOLEA coefficients $\tilde{f}_N$ reported in Ref.~\cite{Zhang:2016pja}. As explained at the end of Sec.~\ref{sec:setup}, the universal coefficients corresponding to operators containing $P$ insertions are not captured by the bosonic UOLEA. Consequently, $g_{10}^i$, $g_{18}^{ij}$ and $g_{19}^{ij}$ will not appear below. To simplify the expressions, we define the sum of two masses as:
\begin{equation*}
m_{ij} \equiv m_i + m_j.
\end{equation*} 
Using this definition, we have:

\begin{align}
g_1^i &= m_i \, \tilde{f}_2^i, \quad \quad g_2^{ij} = \frac{1}{2} \left( \tilde{f}_2^i + \tilde{f}_2^j \right) + m_{ij}^2 \tilde{f}_4^{ij}, \\ 
g_3^{ijk} &= m_{ij} \tilde{f}_4^{ij} +  m_{jk} \tilde{f}_4^{jk} + m_{ki} \tilde{f}_4^{ki} + \frac{3}{2} m_{ij} m_{jk} m_{ki} \tilde{f}_8^{ijk}, \\
g_4^{ijkl} &= \tilde{f}_4^{ik} + \tilde{f}_4^{jl} + \frac{3}{2}  \left[ m_{ij} m_{jk} \tilde{f}_8^{ijk} + m_{jk} m_{kl} \tilde{f}_8^{jkl} + m_{kl} m_{li} \tilde{f}_8^{kli} + m_{li} m_{ij} \tilde{f}_8^{lij}  \right] \notag \\
& + 2 m_{ij} m_{jk} m_{kl} m_{li} \tilde{f}_{10}^{ijkl}, \\[0.2em]
g_5^{ij} &= - \tilde{f}_4^{ij} - m_{ij}^2 \tilde{f}_7^{ij}, \quad \quad g_6^i = - \tilde{f}_3^i + \frac{1}{2} \tilde{f}_4^{ii} \\
g_7^{ijklm} &= \frac{3}{2} \left( m_{ij} \tilde{f}_8^{ijl} + m_{jk} \tilde{f}_8^{jkm} + m_{kl} \tilde{f}_8^{kli} + m_{lm} \tilde{f}_8^{lmj} + m_{mi} \tilde{f}_8^{mik} \right) \notag \\
&+ 2 \left( m_{ij} m_{jk} m_{kl} \tilde{f}_{10}^{ijkl} + m_{jk} m_{kl} m_{lm} \tilde{f}_{10}^{jklm} + m_{kl} m_{lm}  m_{mi}  \tilde{f}_{10}^{klmi} \right. \notag \\ 
& \left. + m_{lm}  m_{mi} m_{ij} \tilde{f}_{10}^{lmij} + m_{mi} m_{ij} m_{jk} \tilde{f}_{10}^{mijk} \right) + \frac{5}{2} m_{ij} m_{jk} m_{kl} m_{lm}  m_{mi} \tilde{f}_{16}^{ijklm} \\
g_8^{ijk} &= -\left( m_{jk} \tilde{f}_7^{jk} + m_{ki} \tilde{f}_7^{ki} \right) - \frac{3}{2} m_{ij} \tilde{f}_8^{ijk} - \frac{1}{2} m_{ij} m_{jk} m_{ki} \tilde{f}_{11}^{ijk} \\
g_9^i &= m_i \left( \frac{3}{2} \tilde{f}_8^{iii} - \tilde{f}_9^i \right), \\
g_{11}^{ijklmn} &= \frac{3}{2} \left( \tilde{f}_8^{ikm} + \tilde{f}_8^{jln} \right) + 2 \left( m_{ij} m_{jk} \tilde{f}_{10}^{ijkm} + m_{jk} m_{kl} \tilde{f}_{10}^{jkln} + m_{kl} m_{lm} \tilde{f}_{10}^{klmi} + m_{lm} m_{mn} \tilde{f}_{10}^{lmnj}  \right. \notag \\ 
&\left. + m_{mn} m_{ni} \tilde{f}_{10}^{mnik} + m_{ni} m_{ij} \tilde{f}_{10}^{nijl} \right) + 2 \left( m_{ij} m_{lm} \tilde{f}_{10}^{ijlm} + m_{jk} m_{mn} \tilde{f}_{10}^{jkmn} + m_{kl} m_{ni} \tilde{f}_{10}^{klni} \right) \notag \\
&+ \frac{5}{2} \left( m_{ij} m_{jk} m_{kl} m_{lm} \tilde{f}_{16}^{ijklm} + m_{jk} m_{kl} m_{lm} m_{mn} \tilde{f}_{16}^{jklmn} + m_{kl} m_{lm} m_{mn} m_{ni} \tilde{f}_{16}^{klmni} \right. \notag \\ 
&\left. + m_{lm} m_{mn} m_{ni} m_{ij} \tilde{f}_{16}^{lmnij} +  m_{mn} m_{ni} m_{ij} m_{jk} \tilde{f}_{16}^{mnijk} +  m_{ni} m_{ij} m_{jk} m_{kl} \tilde{f}_{16}^{nijkl}  \right) \notag \\
&+ 3  m_{ij} m_{jk} m_{kl} m_{lm}  m_{mn} m_{ni} \tilde{f}_{19}^{ijklmn}, \\
g_{12}^{ijkl} &= - \tilde{f}_7^{jl} - \frac{3}{2} \tilde{f}_8^{kli} - 2 m_{ij} m_{jk} \tilde{f}_{10}^{ijkl} - \frac{1}{2} \left( m_{jk} m_{kl} \tilde{f}_{11}^{jkl} + m_{kl} m_{li} \tilde{f}_{11}^{ikl} + m_{li} m_{ij} \tilde{f}_{11}^{ijl}  \right) \notag \\
& + \frac{1}{2} m_{ij} m_{jk} m_{kl} m_{li} \tilde{f}_{17}^{ijkl}, \\
g_{13}^{ijkl} &= - \left( \tilde{f}_7^{ik} + \tilde{f}_7^{jl} \right) - 2 m_{ij} m_{kl} \tilde{f}_{10}^{ijkl} - \frac{1}{2} \left( m_{li} m_{ij} \tilde{f}_{11}^{ijl} + m_{jk} m_{kl} \tilde{f}_{11}^{klj} \right) \notag \\
& - \frac{1}{2} \left( m_{ij} m_{jk} \tilde{f}_{11}^{ijk} + m_{kl} m_{li} \tilde{f}_{11}^{kli}  \right) - m_{ij} m_{jk} m_{kl} m_{li} \tilde{f}_{18}^{ijkl}, \\
g_{14}^{ij} &= \tilde{f}_7^{ij} + m_{ij}^2 \tilde{f}_{12}^{ij}, \\
g_{15}^{ij} &= \tilde{f}_7^{ij} + \frac{3}{4} \left( \tilde{f}_8^{iii} - \tilde{f}_8^{iij} \right) - \frac{1}{2} \tilde{f}_9^i + m_{ij}^2 \left( \tilde{f}_{10}^{iiij} - \frac{1}{2} \tilde{f}_{13}^{ij} - \frac{1}{4} \tilde{f}_{14}^{ij} \right), \\
g_{16}^{ij} &= -2 \tilde{f}_7^{ij} + \frac{3}{4} \left( \tilde{f}_8^{iij} + \tilde{f}_8^{ijj} \right) +  m_{ij}^2 \left( \tilde{f}_{10}^{iijj} + \frac{1}{2} \tilde{f}_{14}^{ij} \right), \\
g_{17}^{ij} &= \tilde{f}_7^{ij} - \frac{3}{2} \tilde{f}_8^{iij} - m_{ij}^2 \left( \frac{1}{2} \tilde{f}_{14}^{ij} + \tilde{f}_{15}^{ij} \right), \\
g_{20}^i &= \tilde{f}_5^i - \tilde{f}_7^{ii}, \quad \quad g_{21}^i = - \frac{3}{2} \tilde{f}_6^i + 3 \tilde{f}_7^{ii} - \frac{3}{2} \tilde{f}_8^{iii}.
\end{align}
 
\FloatBarrier
 
\section{Fermionic Universal Coefficients for Equal Masses} \label{app:equal-masses}
\setcounter{equation}{0}
\renewcommand{\arraystretch}{1.35}

We provide in Tables~\ref{tab:univ-coeff-d-1234}--\ref{tab:univ-coeff-d-6-p2}, listed below, the values of all 56 independent fermionic universal coefficients in the limit where all the VL fermion masses are degenerate. For each coefficient, we provide the corresponding operator and the equation where its full, non--degenerate expression can be found. The full result is obtained by multiplying by $ \frac{c_f n_D}{16 \pi^2} = -\frac{1}{4 \pi^2}$.

\begin{table}[H]
  \centering
  \begin{tabular}{ | c | c | c | c | }
    \hline 
    Operator & Coefficient & \begin{tabular}{@{}c@{}} Equal masses \\ expression \end{tabular} & Eq. \\ \hline \hline
    $S_{ii}$ & $g_1^i$ & $m^3 \left( 1 + \log \frac{\mu^2}{m^2} \right) $ & \eqref{eq:g1} \\ \hline
    $ \frac{1}{2} S_{ij} S_{ji} $ & $g_2^{ij}$ & $m^2 \left( 1 + 3 \log \frac{\mu^2}{m^2} \right) $ & \multirow{2}{*}[-1.6mm]{\eqref{eq:g2}} \\ 
    $ \frac{1}{2} P_{ij} P_{ji} $ & $g_2^{i(j)} + \delta g_2^{ij} $ & $m^2 \left(3 + \log \frac{\mu^2}{m^2} \right) $ &  \\ \hline
    $ \frac{1}{3} S_{ij} S_{jk} S_{ki} $ & $g_3^{ijk}$ & $m  \left( -2 + 3 \log \frac{\mu^2}{m^2} \right) $ & \multirow{2}{*}[-1.6mm]{\eqref{eq:g3}} \\
     $ S_{ij} P_{jk} P_{ki} $ & $g_3^{ij(k)} + \delta g_3^{ij} $ & $ m \left( 2 + \log \frac{\mu^2}{m^2} \right) $ &  \\ \hline
     $ \frac{1}{4} S_{ij} S_{jk} S_{kl} S_{li} $ & $g_4^{ijkl}$ & $ -\frac{8}{3} + \log \frac{\mu^2}{m^2} $ & \multirow{4}{*}[-1.2mm]{\eqref{eq:g4}} \\ 
     $ S_{ij} S_{jk} P_{kl} P_{li}$ & $g_4^{ijk(l)} + \delta g_{4a} $ & $ \log \frac{\mu^2}{m^2} $ & \\
     $ -\frac{1}{2} S_{ij} P_{jk} S_{kl} P_{li} $ & $g_4^{ij(kl)}$ & $\log \frac{\mu^2}{m^2} $ &  \\ 
     $ \frac{1}{4} P_{ij} P_{jk} P_{kl} P_{li} $ & $g_4^{i(j)k(l)} + \delta g_{4b} $ & $ \frac{8}{3} + \log \frac{\mu^2}{m^2} $ & \\ \hline
     $ \frac{1}{2} [D_\mu,S]_{ij} [D^\mu,S]_{ji} $ & $g_5^{ij}$ & $ \frac{1}{3} - \frac{1}{2} \log \frac{\mu^2}{m^2} $ & \multirow{2}{*}[-0.6mm]{(\ref{eq:g5}, \ref{eq:delta-g5})} \\ 
    $ \frac{1}{2} [D_\mu, P]_{ij} [D^\mu, P]_{ji} $ & $g_5^{i(j)} + \delta g_5 $ & $ -\frac{1}{2} -\frac{1}{2}  \log \frac{\mu^2}{m^2}  $ &  \\ \hline
    $ \frac{1}{2} F^{\mu\nu}_i F_{\mu\nu,i}$ & $g_6^i$ & $  \frac{1}{6} \log \frac{\mu^2}{m^2} $ & \eqref{eq:g6} \\ \hline
	\end{tabular}
\caption{Values of the dimension $\leq 4$ fermionic universal coefficients in the limit of equal masses.} \label{tab:univ-coeff-d-1234}
\end{table}

\begin{table}
  \centering
  \begin{tabular}{ | c | c | c | c | }
    \hline 
    Operator & Coefficient & \begin{tabular}{@{}c@{}} Equal masses \\ expression \end{tabular} & Eq. \\ \hline \hline
    $\frac{1}{5} S_{ij} S_{jk} S_{kl} S_{lm} S_{mi}$ & $g_7^{ijklm}$ & $-\frac{1}{2 m}$ & \multirow{4}{*}[-1mm]{\eqref{eq:g7}} \\
     $ S_{ij} S_{jk} S_{kl} P_{lm} P_{mi}$ & $g_7^{ijkl(m)}$ & $-\frac{5}{6 m}$ & \\
     $ S_{ij} S_{jk} P_{kl} S_{lm} P_{mi}$ & $g_7^{ijk(l)(m)}$ & $-\frac{1}{2 m}$ & \\
     $ S_{ij} P_{jk} P_{kl} P_{lm} P_{mi}$ & $g_7^{ij(k)l(m)}$ & $-\frac{1}{2 m}$ & \\ \hline
     $ S_{ij} [D_\mu,S]_{jk} [D^\mu,S]_{ki}$ & $g_8^{ijk}$ & $\frac{1}{2 m}$ & \multirow{4}{*}[-1mm]{\eqref{eq:g8}} \\
     $ S_{ij} [D_\mu,P]_{jk} [D^\mu,P]_{ki}$ & $g_8^{ij(k)}$ & $\frac{1}{2 m}$ & \\
     $ P_{ij} [D_\mu,P]_{jk} [D^\mu,S]_{ki}$ & $g_8^{i(j)k}$ & $\frac{1}{6 m}$ & \\
     $ P_{ij} [D_\mu,S]_{jk} [D^\mu,P]_{ki}$ & $g_8^{(i)jk}$ & $\frac{1}{6 m}$ & \\ \hline
      $S_{ii} F^{\mu\nu}_i F_{\mu\nu,i} $ & $g_9^i $ & $- \frac{1}{6 m}$ & \eqref{eq:g9} \\ \hline
      $P_{ii} \widetilde F^{\mu\nu}_i F_{\mu\nu,i} $ & $g_{10}^i $ & $ \frac{1}{4 m} $ & \eqref{eq:g10} \\ \hline
	\end{tabular}
\caption{Values of the dimension--5 fermionic universal coefficients in the limit of equal masses.} \label{tab:univ-coeff-d-5}
\end{table}

\begin{table}
  \centering
  \begin{tabular}{ | c | c | c | c | }
    \hline 
    Operator & Coefficient & \begin{tabular}{@{}c@{}} Equal masses \\ expression \end{tabular} & Eq. \\ \hline \hline
  $\frac{1}{6}  S_{ij} S_{jk} S_{kl} S_{lm} S_{mn} S_{ni}$ & $g_{11}^{ijklmn}$ & $\frac{1}{10 m^2}$ & \multirow{8}{*}[-0.3mm]{\eqref{eq:g11}} \\
  $ S_{ij} S_{jk} S_{kl} S_{lm} P_{mn} P_{ni}$ & $g_{11}^{ijklm(n)}$ & $\frac{1}{6 m^2}$ & \\
  $ - S_{ij} S_{jk} S_{kl} P_{lm} S_{mn} P_{ni}$ & $g_{11}^{ijkl(m)(n)}$ & $-\frac{1}{6 m^2}$  & \\
  $ \frac{1}{2} S_{ij} S_{jk} P_{kl} S_{lm} S_{mn} P_{ni}$ & $g_{11}^{ijk(l)(m)(n)}$ & $ -\frac{1}{2 m^2}$ & \\
   $ S_{ij} S_{jk} P_{kl} P_{lm} P_{mn} P_{ni}$  & $g_{11}^{ijk(l)m(n)}$  & $-\frac{1}{6 m^2}$ & \\
   $ -S_{ij} P_{jk} S_{kl} P_{lm} P_{mn} P_{ni}$ & $g_{11}^{ij(k)(l)m(n)}$ & $-\frac{1}{2 m^2}$ & \\
   $ \frac{1}{2} S_{ij} P_{jk} P_{kl} S_{lm} P_{mn} P_{ni}$ & $g_{11}^{ij(k)lm(n)}$ & $ - \frac{1}{6 m^2}$ & \\
   $ \frac{1}{6} P_{ij} P_{jk} P_{kl} P_{lm} P_{mn} P_{ni}$ & $g_{11}^{i(j)k(l)m(n)}$ & $-\frac{1}{2 m^2}$ & \\ \hline
   $ S_{ij} S_{jk} [D_\mu , S]_{kl} [D^\mu , S]_{li} $ & $g_{12}^{ijkl}$ & $ -\frac{1}{10 m^2}$ & \multirow{8}{*}[-0.3mm]{\eqref{eq:g12}} \\
  $ S_{ij} S_{jk} [D_\mu , P]_{kl} [D^\mu , P]_{li} $ & $g_{12}^{ijk(l)}$ & $0$ & \\
  $ S_{ij} P_{jk} [D_\mu , P]_{kl} [D^\mu , S]_{li} $ & $g_{12}^{ij(k)l}$ & $\frac{1}{6 m^2}$  & \\
  $ P_{ij} P_{jk} [D_\mu , S]_{kl} [D^\mu , S]_{li} $ & $g_{12}^{i(j)kl}$ &  $\frac{1}{6 m^2}$ & \\
   $ P_{ij} S_{jk} [D_\mu , S]_{kl} [D^\mu , P]_{li} $  & $g_{12}^{(i)jkl}$  &  $\frac{1}{6 m^2}$ & \\
   $ -S_{ij} P_{jk} [D_\mu , S]_{kl} [D^\mu , P]_{li} $ & $g_{12}^{ij(k)(l)}$ & $\frac{1}{3 m^2}$ & \\
   $ -P_{ij} S_{jk} [D_\mu , P]_{kl} [D^\mu , S]_{li} $ & $g_{12}^{i(j)(k)l}$ & $\frac{1}{3 m^2}$ & \\
   $ P_{ij} P_{jk} [D_\mu , P]_{kl} [D^\mu , P]_{li} $ & $g_{12}^{i(j)k(l)}$ & $\frac{1}{3 m^2}$ & \\ \hline
   $ \frac{1}{2} S_{ij} [D_\mu , S]_{jk} S_{kl}  [D^\mu , S]_{li} $ & $g_{13}^{ijkl}$ & $ -\frac{3}{10 m^2}$ & \multirow{6}{*}[-0.3mm]{\eqref{eq:g13}} \\
   $ S_{ij} [D_\mu , S]_{jk} P_{kl}  [D^\mu , P]_{li} $ & $g_{13}^{ijk(l)}$ & $ 0 $ & \\
   $ S_{ij} [D_\mu , P]_{jk} P_{kl}  [D^\mu , S]_{li} $ & $g_{13}^{ij(k)l}$ & $ 0 $ & \\ 
   $ -\frac{1}{2} S_{ij} [D_\mu , P]_{jk} S_{kl}  [D^\mu , P]_{li} $ & $g_{13}^{ij(k)(l)}$ & $ \frac{1}{2 m^2}$ & \\
   $  -\frac{1}{2} P_{ij} [D_\mu , S]_{jk} P_{kl}  [D^\mu , S]_{li} $ & $g_{13}^{i(j)(k)l}$ & $ \frac{1}{6 m^2}$ & \\
   $ \frac{1}{2} P_{ij} [D_\mu , P]_{jk} P_{kl}  [D^\mu , P]_{li} $ & $g_{13}^{i(j)k(l)}$ & $ \frac{1}{6 m^2}$ & \\ \hline
	\end{tabular}
\caption{Values of the $\mathcal{O}(X^6)$ and $\mathcal{O}(X^4 D^2)$ fermionic universal coefficients in the limit of equal masses.} \label{tab:univ-coeff-d-6-p1}
\end{table}

\begin{table}
  \centering
  \begin{tabular}{ | c | c | c | c | }
    \hline 
    Operator & Coefficient & \begin{tabular}{@{}c@{}} Equal masses \\ expression \end{tabular} & Eq. \\ \hline \hline
 $ \frac{1}{2} [D_\mu ,[ D^\mu ,S]]_{ij} [D_\nu ,[ D^\nu ,S]]_{ji} $ & $g_{14}^{ij}$ & $ - \frac{1}{20 m^2}$  & \multirow{2}{*}[-1.6mm]{\eqref{eq:g14}} \\ 
    $ \frac{1}{2} [D_\mu ,[ D^\mu ,P]]_{ij} [D_\nu ,[ D^\nu ,P]]_{ji} $ & $g_{14}^{i(j)}$ & $ - \frac{1}{12 m^2}$  &  \\ \hline
   $ S_{ij} S_{ji} F^{\mu\nu}_i F_{\mu\nu,i} $ & $g_{15}^{ij}$ & $ \frac{7}{120 m^2} $ & \multirow{2}{*}[-1.6mm]{\eqref{eq:g15}} \\
   $ P_{ij} P_{ji} F^{\mu\nu}_i F_{\mu\nu,i} $ & $g_{15}^{i(j)}$ & $ -\frac{1}{24 m^2} $ &  \\ \hline
   $ \frac{1}{2} S_{ij}  F^{\mu\nu}_j S_{ji} F_{\mu\nu,i} $ & $g_{16}^{ij}$ & $ \frac{1}{20 m^2}$ & \multirow{2}{*}[-1.6mm]{\eqref{eq:g16}} \\
    $ \frac{1}{2} P_{ij}  F^{\mu\nu}_j P_{ji} F_{\mu\nu,i} $ & $g_{16}^{i(j)}$ & $ - \frac{1}{12 m^2}$ &  \\ \hline
   $ i \, S_{ij} [D_\mu, S]_{ji} [D_\nu, F^{\nu\mu}]_i  $ & $g_{17}^{ij}$ & $\frac{2}{15 m^2}$ & \multirow{2}{*}[-1.6mm]{\eqref{eq:g17}} \\
    $ i \, P_{ij} [D_\mu, P]_{ji} [D_\nu, F^{\nu\mu}]_i  $ & $g_{17}^{i(j)}$ &  $\frac{1}{6 m^2}$ &  \\ \hline
    $ ( S_{ij} P_{ji} + P_{ij} S_{ji} ) \widetilde F^{\mu\nu}_i F_{\mu\nu,i} $ & $g_{18}^{i(j)}$ & $ -\frac{1}{12 m^2} $ & \eqref{eq:g18+g19} \\ \hline
    $ S_{ij}  \widetilde F^{\mu\nu}_j P_{ji} F_{\mu\nu,i} $ & $g_{19}^{i(j)}$ & $ -\frac{1}{12 m^2} $ & \eqref{eq:g18+g19} \\ \hline
   $ \frac{1}{2}  \left[ D^\mu  , F_{\mu\nu} \right]_i \left[ D_\rho  , F^{\rho\nu} \right]_i $ & $g_{20}^i$ & $\frac{1}{15 m^2}$  & \eqref{eq:g20+g21} \\ \hline
   $ \frac{i}{3} F^{\mu}_{\;\;\: \nu} F^{\nu}_{\;\;\: \rho} F^{\rho}_{\;\;\: \mu} $ & $g_{21}^i$ & $\frac{1}{60 m^2}$  & \eqref{eq:g20+g21}  \\ \hline
	\end{tabular}
\caption{Values of the $\mathcal{O}(X^2 D^4)$, $\mathcal{O}( S P D^4)$, and $\mathcal{O}(D^6)$ fermionic universal coefficients in the limit of equal masses.} \label{tab:univ-coeff-d-6-p2}
\end{table}

\FloatBarrier

\newpage

\bibliographystyle{utphys}
\bibliography{vlf-dim6-ref}

\end{document}